\documentclass[prx,aps,twocolumn,amsfonts,showpacs,nofootinbib,preprintnumbers,superscriptaddress]{revtex4-1}

\usepackage{multibib}
\usepackage{colortbl}
\usepackage{url}
\usepackage{hyperref}

\usepackage{amsmath,amsthm,amssymb} 
\usepackage{dsfont,yfonts}
\usepackage{hyperref}
\usepackage{floatrow,dcolumn,rotating}
\usepackage{array,xcolor,graphicx}
\usepackage{graphics}
\usepackage{graphicx}
\usepackage{soul}

\hypersetup{colorlinks=true,linkcolor=blue,citecolor=red,urlcolor=blue}

\begin{document}

\title{Enhanced Gravity Model of trade:\\ reconciling macroeconomic and network models}
\author{Assaf Almog}
\affiliation{Tel Aviv University, Department of Industrial Engineering, 69978 Tel Aviv, Israel}
\author{Rhys Bird}
\affiliation{Instituut-Lorentz for Theoretical Physics, Leiden University, Niels Bohrweg 2,  2333 CA Leiden, Netherlands}
\author{Diego Garlaschelli}
\affiliation{IMT School for Advanced Studies, Piazza S. Francesco 19, 55100 Lucca, Italy }
\affiliation{Instituut-Lorentz for Theoretical Physics, Leiden University, Niels Bohrweg 2,  2333 CA Leiden, Netherlands}


\begin{abstract}
The structure of the International Trade Network (ITN), whose nodes and links represent world countries and their trade relations respectively, affects key economic processes worldwide, including globalization, economic integration, industrial production, and the propagation of shocks and instabilities. Characterizing the ITN via a simple yet accurate model is an open problem. The traditional Gravity Model (GM) successfully reproduces the volume of trade between connected countries, using macroeconomic properties such as GDP, geographic distance, and possibly other factors. However, it predicts a network with complete or homogeneous topology, thus failing to reproduce the highly heterogeneous structure of the ITN. On the other hand, recent maximum-entropy network models successfully reproduce the complex topology of the ITN, but provide no information about trade volumes. Here we integrate these two currently incompatible approaches via the introduction of an Enhanced Gravity Model (EGM) of trade. The EGM is the simplest model combining the GM with the network approach within a maximum-entropy framework. Via a unified and principled mechanism that is transparent enough to be generalized to any economic network, the EGM provides a new econometric framework wherein trade probabilities and trade volumes can be separately controlled by any combination of dyadic and country-specific macroeconomic variables. The model successfully reproduces both the global topology and the local link weights of the ITN, parsimoniously reconciling the conflicting approaches. It also indicates that the probability that any two countries trade a certain volume should follow a geometric or exponential distribution with an additional point mass at zero volume.\end{abstract}

\maketitle

\section{Introduction}
The International Trade Network (ITN) is the complex network of trade relationships existing between pairs of countries in the world. The nodes (or vertices) of the ITN represent nations and the edges (or links) represent their (weighted) trade connections. 
In a global economy extending across national borders, there is increasing interest in understanding the mechanisms involved in trade interactions and how the position of a country within the ITN may affect its economic growth and integration \cite{economicnetworks,Kali1,Kali2,integration,Spitz}. Moreover, in the wake of recent financial crises the interconnectedness of economies has become a matter of concern as a source of instability~\cite{myscience}. 
As the modern architecture of industrial production extends over multiple countries via geographically wider supply chains, sudden changes in the exports of a country (due e.g. to unpredictable financial, environmental, technological or even political circumstances) can rapidly propagate to other countries via the ITN. The assessment of the associated trade risks requires detailed information about the underlying network structure \cite{TradeRisk}.
In general, among the possible channels of interaction among countries, trade plays a major role \cite{Kali1,Kali2,integration}. 

The above considerations imply that the  empirical structure of the ITN plays a crucial role in increasingly many economic phenomena of global relevance. 
It is therefore becoming more and more important to characterize the ITN via simple but accurate models that identify both the basic ingredients and the mathematical expressions required to accurately reproduce the details of the empirical network structure.
Reliable models of the ITN can better inform economic theory, foreign policy, and the assessment of trade risks and instabilities worldwide.

In this paper, we emphasize that current models of the ITN have strong limitations, and that none of them is satisfactory, from either a theoretical or a phenomenological point of view. 
We point out equally strong (and largely complementary) problems affecting on one hand traditional macroeconomic models, which focus on the local \emph{weight} of the links of the network, and on the other hand more recent network models, which focus on the \emph{existence} of links, i.e. on the global topology of the ITN. 
We then introduce a new model of the ITN that preserves all the good ingredients of the models proposed so far, while at the same time improving upon the limitations of each of them.
The model can be easily generalized to any (economic) network and provides an explicit specification of the full probability distribution that a given pair of countries is connected by a certain volume of trade, fixing an otherwise arbitrary choice in previous approaches.
This distribution is found to be either geometric (for discrete volumes) or exponential (for continuous volumes), with an additional point mass at zero volume. 
This feature, which is different from all previous specifications of international trade models, is shown to replicate both the local trade volumes and the global topology of the empirical ITN remarkably well. 

\section{Preliminaries: building blocks of the model}
Before we fully specify our model, we preliminarily identify its building blocks by reviewing the strengths and weaknesses of the two main modelling frameworks adopted so far. 

\subsection{Gravity models of trade\label{sec:gravity}}
We start by discussing traditional macroeconomic models of international trade. 
These models have mainly focused on the \emph{volume} (i.e. the value e.g. in dollars) of trade between countries, largely because the economic literature perceives trade volumes as being \emph{a priori} more informative than the topology of the ITN: the striking heterogeneity of trade volumes observed between different pairs of countries is clearly not captured by a purely `binary' description where all connections are effectively given the same weight. 
Based on this argument, emphasis has been put on explaining the (expected) volume of trade between two countries, given certain dyadic and country-specific macroeconomic properties.

Jan Tinbergen, the physics-educated\footnote{Jan Tinbergen studied physics in Leiden, where he carried out a PhD under the supervision of the theoretical physicist Paul Ehrenfest. Tinbergen defended his thesis in 1929, and then became a leading economist. He was awarded the first Nobel memorial prize in economics in 1969.} Dutch  economist who was awarded the first Nobel memorial prize in economics, introduced the so-called Gravity Model (GM) of trade \cite{Tinbergen1}.
The GM aims at inferring the volume of trade from the knowledge of Gross Domestic Product, mutual geographic distance, and possibly additional dyadic factors of macroeconomic relevance \cite{gravity_book,DeBene2}.
In one of its simplest forms, the GM predicts that, if $i$ and $j$ label two different countries ($i,j=1,\dots,N$ where $N$ is the total number of countries in the world), then the expected volume of trade from $i$ to $j$ is
\begin{equation}
\langle w_{ij}\rangle= c~\textrm{GDP}^{\alpha}_i~\textrm{GDP}^{\beta}_j~R^{-\gamma}_{ij}\qquad c,\alpha,\beta,\gamma>0,
\label{eq_gravity}
\end{equation}
where $\textrm{GDP}_k$ is the Gross Domestic Product of country $k$, $R_{ij}$ is the geographic distance between countries $i$ and $j$, and $c,\alpha,\beta,\gamma$ are free global parameters to be estimated.
In the above \emph{directed} specification of the GM, the flows $w_{ij}$ and $w_{ji}$ can be different. 
An analogous \emph{undirected} specification exists, where the volumes of trade from $i$ to $j$ and from $j$ to $i$ are added together into a single value $w_{ij}=w_{ji}$ of bilateral trade. In the latter case, Eq.~\eqref{eq_gravity} still holds but with the symmetric choice $\alpha=\beta$.
With this in mind, we will keep our discussion entirely general throughout the paper and, unless otherwise specified, allow all quantities to be interpreted either as directed or as undirected. 
Only in our final empirical analysis we will adopt an undirected description for simplicity.

More complicated variants of Eq.~\eqref{eq_gravity} use additional factors (with associated free parameters) either favoring or resisting trade~\cite{gravity_book,DeBene2}.
Like the GDP and geographic distances, these factors can be either country-specific (e.g. population) or dyadic (e.g. common currency, trade agreements, shared borders, common language, etc.).
In general, if we collectively denote with $\vec{n}_i$ the vector of all \emph{node-specific} factors and with $\vec{D}_{ij}$ the vector of all \emph{dyad-specific} factors used (note that these vectors may have different dimensionality), Eq.~\eqref{eq_gravity} can be generalized to
\begin{equation}
\langle w_{ij}\rangle= F_{\vec{\phi}}\,(\vec{n}_i,\vec{n}_j,\vec{D}_{ij})\qquad F> 0,
\label{eq_gravityF}
\end{equation}
where the functional form of $F_{\vec{\phi}}\,(\vec{n}_i,\vec{n}_j,\vec{D}_{ij})$ need not be of the same type as in Eq.~\eqref{eq_gravity}, and ${\vec{\phi}}$ is a vector containing all the free parameters of the model (like $c,\alpha, \beta, \gamma$ for the particular case above).
Indeed, although in this paper we focus on the GM applied to the international trade network, our discussion equally applies to many other models of (socio-economic) networks as well. 
For instance, the recently proposed Radiation Model (RM)~\cite{RMmodel} is also described by Eq.~\eqref{eq_gravityF}, where $\vec{n}_i$ and $\vec{D}_{ij}$ include certain geographical and demographical variables. 
Our following discussion applies to both the GM and the RM, as well as any other model described by Eq.~\eqref{eq_gravityF}.
Similarly, it does not only apply to trade networks, since both the GM and the RM have been successfully applied to other systems as well, including mobility and traffic flows~\cite{RMmodel,zipf,balcan,koreaGM}, communication networks~\cite{LambiotteGM}, and
migration patterns~\cite{migrationGM} (the latter representing - to our knowledge - the earliest application of the GM to a socio-economic system, dating back to 1889~\cite{ravenstein}).

 \begin{figure*}[t]
  \centering
  \includegraphics[width=.9\linewidth]{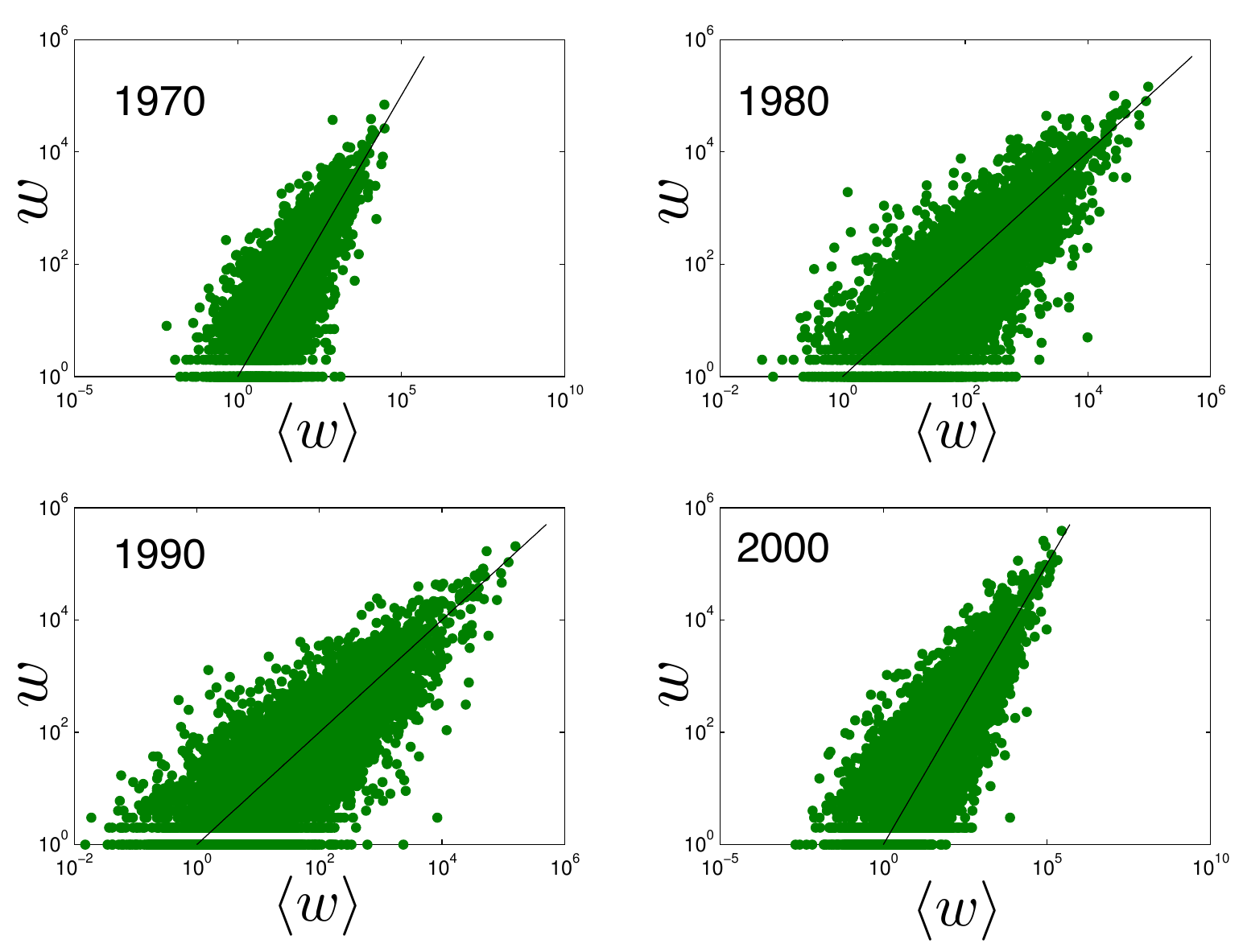}
\caption{{\bf Empirical non-zero trade flows vs. the corresponding expectation under the traditional Gravity Model.} Log-log plot comparing the empirical volume ($y$-axis) of all non-zero bilateral trade flows in the ITN with the corresponding expected volume ($x$-axis) predicted by the Gravity Model defined in Eq.~\eqref{eq_gravity}, with parameters estimated as reported in Table~\ref{parameters-1}. Top left: year 1970, top right: year 1980, bottom left: year 1990, bottom right: year 2000.
The black line is the identity line corresponding to the ideal, perfect match that would be achieved if the empirical weights were exactly equal to their expected values, i.e. in complete absence of randomness.}
\label{fig-GMflows}
\end{figure*}

It is generally accepted that the expected trade volumes postulated by the GM, already in its simplest form given by Eq.~\eqref{eq_gravity}, are in good agreement with the observed flows between trading countries. 
To illustrate this result, in Fig.~\ref{fig-GMflows} we show a typical log-log plot comparing the empirical volume of the realized (bilateral) international trade flows with the corresponding expected values calculated under the GM as defined in Eq.~\eqref{eq_gravity} (with parameters calculated as reported in Table~\ref{parameters-1}).
The figure shows the typical qualitative consistency between the GM and the empirical non-zero trade volumes.
However, it should be noted that, while Eqs.~\eqref{eq_gravity} and~\eqref{eq_gravityF} define the expected value of $w_{ij}$, the full probability distribution from which this expected value is calculated is not specified, and actually depends on how the model is implemented in practice. In the GM case, the distribution is chosen to be either Gaussian (corresponding to additive noise, in which case the expected weights can be fitted to the observed ones via a simple linear regression \cite{Glick&Rose_2001,Rose&Spiegel_2004}), log-normal (corresponding to multiplicative noise and requiring a linear regression of log-transformed weights~\cite{SantosSilva&Tenreyro_2006} as we did to produce Fig.~\ref{fig-GMflows} and Table~\ref{parameters-1}), Poisson \cite{SantosSilva&Tenreyro_2006}, or  more sophisticated \cite{Linders&deGroot_2006} (see \cite{Giorgio_gravity} for a review). The arbitrariness of the weight distribution already highlights a fundamental weakness of the traditional formulation of the model. 
Moreover, for both additive and multiplicative Gaussian noise, the model can produce undesired negative values.

\begin{table}[b]
\centering
\begin{tabular}{|c|c|c|c|}
\hline
\multicolumn{4}{|c|}{\bf Traditional Gravity Model} \\ \hline\hline
Year & c & $\alpha$, $\beta$ & $\gamma$ \\
\hline \hline
1970                     & $9.9\cdot10^{8}$ &0.91        & 0.81     \\ \hline 
1980                     &  $3.1\cdot10^{9}$ & 0.83      & 0.89     \\ \hline
1990                     &  $1.5\cdot10^{10}$ & 0.97     & 0.93     \\ \hline
2000                     &  $4.3\cdot10^{10}$  & 1.05    &  0.93     \\ \hline
\end{tabular}
\caption{Parameter values for the traditional Gravity Model used in Fig.~\ref{fig-GMflows} and calculated by fitting Eq.~\eqref{eq_gravity} (with the symmetry constraint $\alpha\equiv\beta$) to all non-zero empirical bilateral trade flows via an Ordinary Least Square (OLS) regression of log-transformed weights.}
\label{parameters-1}
\end{table}

A related but more fundamental limitation of the GM is that, at least in its simplest and most natural implementations, it cannot generate zero volumes -- thereby predicting a fully connected network~\cite{Giorgio_gravity0,Giorgio_gravity,Tinbergen2}. 
This means that the GM can be fitted only to the \emph{non-zero} weights, i.e. the volumes existing between pairs of \emph{connected} countries. 
If used in this way, the model effectively disregards the empirical structure of the network, both as input (thus making predictions on the basis of incomplete data) and as output (thus failing to reproduce the topology).
Operatively, the GM can be used only {\em after} the presence of a trade link has been established independently \cite{Giorgio_gravity}. 
As observed in~\cite{Linders&deGroot_2006},
\emph{``Omitting zero-flow observations implies that we loose information on the causes of (very) low trade''}, because any fit to positive-only flows would significanlty underestimate the effects of factors that diminish trade.
This problem is particularly critical since roughly half of the possible links are found \emph{not} to be realized in the real ITN \cite{fitness,Squartini-random1,Squartini-random2,Squartini3}.
Clearly, the same problem holds for the RM and any more general model of the form specified in Eq.~\eqref{eq_gravityF}.

While there are variants and extensions of the GM that do generate zero weights and a realistic link density (e.g. the so-called Poisson pseudo-maximum likelihood
models \cite{SantosSilva&Tenreyro_2006} and `zero-inflated' gravity models \cite{Linders&deGroot_2006}), these variants systematically fail in reproducing the observed topology \cite{Giorgio_gravity,DeBene2}. 
In other words, while these models can generate the correct number of connections, they tend to put many of the latter in the `wrong place' in the network.
Indeed, even in its generalized forms, the GM predicts a largely homogeneous network structure, while the empirical topology of the ITN is much more heterogeneous and complex \cite{Giorgio_gravity0,Giorgio_gravity}. 
Established empirical signatures of this heterogeneity include a broad distribution of the degree (number of connections) and the strength (total trade volume) of countries~\cite{fitness,Squartini-random1,Squartini-random2,Squartini3,Serrano,Garlaschelli,Vespignani,Caldarelli,Fagiolo1,Fagiolo2,Fagiolo3}, the rich-club phenomenon (whereby well-connected countries are also connected to each other)~\cite{Colizza,Zlatic}, strong clustering and (dis)assortative patterns~\cite{Squartini-random1,Squartini-random2}. 
These highly skewed structural properties are remarkably stable over time. However, their are not replicated by any current version of the GM~\cite{Giorgio_gravity}.

\subsection{Network models of trade\label{sec:network}}
As we mentioned at the beginning, many processes of great economic relevance crucially depend on the large-scale topology of the ITN. 
In light of this result, the sharp contrast between the observed topological complexity of the ITN and the homogeneity of the network structure generated by the GM (including its extensions) calls for major improvements in the modelling approach. 
In particular, in assessing the performance of a model of the ITN, emphasis should be put on how reliably the (global) empirical network structure, besides the (local) volume of trade, is replicated.
In the network science literature, successful models of the ITN have been derived from the \emph{Maximum Entropy Principle}~\cite{fitness,Squartini-random1, Squartini-random2,Squartini3,Tinbergen2,Unbiased,fronczakME,ECM,ECM2,Almog,mybook,mynaturereviews}. 
These models construct ensembles of random networks that have some desired topological property (taken as input from empirical data) and are maximally random otherwise. 
Typically, the constrained properties are chosen to be the degrees and/or the strenghts of all nodes.
In this way the models can perfectly replicate the observed strong heterogeneity of these purely local properties, and at the same time illustrate its immediate (i.e. prior to invoking any other more complicated network formation mechanism) structural effects on any higher-order topological property of the network.
In the different context of financial networks, where the main challenge is a reliable inference of the unobserved topology of a network (typically of interconnected firms or banks) starting from partial, node-aggregate information~\cite{myphysicsreports}, maximum-entropy models have recently turned out to deliver the best-performing reconstruction methods so far~\cite{mybook,mynaturereviews,myphysicsreports}.

In general, different choices of the constrained properties lead to different degrees of agreement between the model and the data.
This can generate intriguing and counter-intuitive insight about the structure of the ITN.
For instance, contrary to what naive economic reasoning would predict, it turns out that the knowledge of purely binary local properties (e.g. node degrees) can be \emph{more informative} than the knowledge of the corresponding weighted properties (e.g. node strengths). 
Indeed, while the binary network reconstructed only from the knowledge of the degrees of all countries is found to be topologically very similar to the real ITN, the weighted network reconstructed only from the strengths of all countries is found to be much denser and very different from the real network \cite{Squartini-random1,Squartini-random2,Squartini3}. 
This is somewhat surprising, given that the economic literature largely postulates that weighted properties are \emph{per se} more informative than the corresponding binary ones.

The solution to this apparent paradox lies in the fact that, while the knowledge of the \emph{entire} weighted network is necessarily more informative than that of its binary projection (in accordance with economic postulates), the knowledge of certain \emph{marginal} properties of the weighted network can be unexpectedly less informative than the knowledge of the corresponding marginal properties of the binary network.
In fact, it turns out that if the degrees of countries are (not) specified in addition to the strengths of countries, the resulting maximum-entropy model can(not) reproduce the empirical weighted network of international trade satisfactorily~\cite{Squartini-random2,ECM,ECM2}. 

An important take-home message is that, in contrast with the mainstream literature, models of the ITN should aim at reproducing not only the strength of countries (as the GM automatically does by approximately reproducing all non-zero weights), but also their degree (i.e. the number of trade partners) \cite{Squartini-random1,Squartini-random2,Squartini3,ECM2}.
In addition to these studies, an alternative approach, the Linear Gravity Transportation Model (LGTM), has also demostrated the importance of the ITN topology~\cite{LGTM}. In this model the monetary flow is balanced for each country (node) based on the number of trade partners (degree). The model produces expectations of the GDP of countries that are consistent with real data, using both the volume of trade flows and the topology of ITN as input.
These studies indicate that, in order to devise improved models of the ITN, one should include the degrees, which are purely topological properties, among the main target quantities to replicate. This is the guideline we will follow in this paper.

Unlike the GM, maximum-entropy models of trade are \emph{a priori} non-explanatory, i.e. they take as input structural properties (as opposed to explanatory economic factors) to explain other structural properties. 
However, they can in fact be used to select \emph{a posteriori} an explicit, empirically validated functional dependence of the structure of the ITN on underlying explanatory factors.
For models with country-specific constraints, this operation can be carried out as follows.
Mathematically, controlling for node-specific properties  is realized by assigning one or more Lagrange multipliers, also known as `hidden variables' or `fitness parameters' $\vec{x}_i$, to each node.
If a certain choice of local constraints is found to replicate the higher-order properties of the real-world network satisfactorily, then one can look for an empirical relationship between the values of the associated hidden variables and those of candidate non-topological, country-specific factors of the type $\vec{n}_i$, like the GDP or total import/export. 
If the hidden variables are indeed (at least approximately) found to be functions of some country-specific factors (i.e. if $\vec{x}_i\approx \vec{f}(\vec{n}_i)$), then one can replace $\vec{x}_i$ with $\vec{f}(\vec{n}_i)$ in the maximum-entropy model, thus reformulating the latter as a model with explanatory variables (i.e. `regressors') of trade, precisely like the GM.
Already in one of the earliest studies on the ITN topology~\cite{Garlaschelli}, the approach outlined above led to the definition of a GDP-driven model for the binary structure of the network, where $\vec{x}_i\propto \textrm{GDP}_i$ (i.e., in this case $\vec{x}_i$ is taken to be one-dimensional). 
The model, which is a reformulation of a maximum-entropy model for binary networks with given degrees, predicts that the probability of a trade connection existing from country $i$ to country $j$ is
\begin{equation}
p_{ij}=\frac{\delta~\textrm{GDP}_i~\textrm{GDP}_j}{1+\delta~\textrm{GDP}_i~\textrm{GDP}_j}\qquad\delta>0,
\label{eq:fermi}
\end{equation}
where $\delta$ is a free parameter that allows to reproduce the empirical link density. 
The model has been tested successfully in multiple ways~\cite{Tinbergen2,fitness,Garlaschelli,Caldarelli,Unbiased}.

The GM in Eq.~\eqref{eq_gravity} and the maximum-entropy model in Eq.~\eqref{eq:fermi} have complementary strengths and weaknesses, the former being a good model for non-zero volumes (while being a bad model for the topology) and the latter being a good model for the topology (while providing no information about trade volumes).
An attempt to reconcile these two complementary and currently incompatible approaches has been recently proposed via the definition of an extension of the maximum-entropy model to the case of weighted networks~\cite{Almog}.
Since, as we mentioned, a maximum-entropy model of weighted networks with given strengths and degrees~\cite{ECM} can correctly replicate many structural properties of the ITN~\cite{ECM2}, it makes sense to reformulate such model as an economically inspired model of the ITN.
Indeed, like in the binary case, the hidden variables enforcing the constraints are found to be strongly correlated with the GDP, thus allowing to express both $p_{ij}$ and $\langle w_{ij}\rangle$ as functions of the GDP~\cite{Almog}. The resulting model is confirmed to be in good accordance with both the topology and the volumes observed in the real ITN.

Unfortunately, in the above approach the choice of country-specific constraints (degrees and strengths) only allows for regressors that have a corresponding country-specific nature. This makes the model in Ref.~\cite{Almog} incompatible with the inclusion of dyadic variables of the type $\vec{D}_{ij}$ and represents a strong limitation for (at least) two reasons. 
Firstly, one of the main lessons learnt from the traditional GM is that the addition of geographic distances improves the fit to the empirical volumes significantly. 
Indeed, in the light of the large body of knowledge accumulated in the international economics literature, it is hard to imagine a realistic and economically meaningful model of international trade that does not allow for simple pair-wise quantities controlling for trade \emph{costs} and \emph{incentives}, including geography~\cite{gravity_book,DeBene2}.
Secondly, even if the structure of the ITN can be replicated satisfactorily in terms of the  `GDP-only' model defined in Eq.~\eqref{eq:fermi}~\cite{fitness,Garlaschelli,Caldarelli}, recent analyses have found evidence that certain metric (although not necessarily geographic\footnote{Building on the hypothesis of the existence of underlying hidden metric spaces in which real-world networks are embedded, Ref.~\cite{worldtradeatlas} models the ITN by looking for an optimal embedding of countries in some abstract metric space. The resulting inferred distances are interpreted as incorporating all possible sources of empirically revealed trade costs, possibly including geographic distances as well. However, since the postulated embedding space is either a unidimensional circle or a hyperbolic plane, these distances are necessarily different from the usual geographic distances $R_{ij}$ appearing in the GM and measured as geodesics on our spherical tridimensional world.}) distances do also play a role in determining the topology of the ITN~\cite{worldtradeatlas}.
Together, these two pieces of evidence call for an inclusion of dyadic factors in $\langle w_{ij}\rangle$ and $p_{ij}$, and highlight a limitation of current maximum-entropy models based only on country-specific constraints.

Combining all the above considerations, it is clear that an improved model of the ITN should aim at retaining the realistic trade volumes postulated by models based on Eq.~\eqref{eq_gravityF} (including the GM, the RM, and possibly many more), while combining them with a realistic network topology generated by (extensions of) maximum-entropy models.
Such a model should also aim at providing the full probability distribution, and not only the expected values as in Eq.~\eqref{eq_gravity}, of trade flows and, unlike the GDP-only model in Eq.~\eqref{eq:fermi}~\cite{fitness} or its current weighted extension~\cite{Almog}, allow for the inclusion of both dyadic and node-specific macroeconomic factors.

\section{The Enhanced Gravity Model of international trade}
In this Section, we introduce what we call the Enhanced Gravity Model (EGM) of trade. 
The EGM mathematically formalizes the two ingredients that, in the light of the previous discussion, any `good' model of economic networks should feature: namely, realistic (trade) volumes and a realistic topology, both controllable by macroeconomic factors.

\subsection{A single model for topology and weights}
The first lesson we have learnt is that Eq.~\eqref{eq_gravityF} is successful in reproducing link weights only \emph{after} the existence of the links themselves has been preliminarly established. 
This implies that Eq.~\eqref{eq_gravityF}, as a model of real-world trade flows, is actually unsatisfactory and should rather be reformulated as a \emph{conditional expectation} of the weight $w_{ij}$, given that $w_{ij}>0$. 
In other words, if $a_{ij}$ denotes the entry of the adjacency matrix $\mathbf{A}=\Theta(\mathbf{W})$ of the ITN (defined via the step function as $a_{ij}=\Theta(w_{ij})$, i.e. $a_{ij}=1$ if $w_{ij}>0$ and $a_{ij}=0$ if $w_{ij}=0$), an improved model should be such that Eq.~\eqref{eq_gravityF} is replaced by
\begin{equation}
\langle w_{ij}|a_{ij}=1\rangle= F_{\vec{\phi}}\,(\vec{n}_i,\vec{n}_j,\vec{D}_{ij})\qquad F>0,
\label{eq_gravityF1}
\end{equation}
where $\langle w_{ij}|a_{ij}=1\rangle$ is the \emph{conditional expected weight} of the trade link from country $i$ to country $j$, \emph{given that such link exists}.
This operation ensures that, whatever the new model looks like, its predictions for the expected trade volume between connected pairs of countries remain identical to the ones proposed in more traditional macroeconomic models.
For instance, choosing $F_{\vec{\phi}}\,(\vec{n}_i,\vec{n}_j,\vec{D}_{ij})=c~\textrm{GDP}^{\alpha}_i~\textrm{GDP}^{\beta}_j~R^{-\gamma}_{ij}$ as in Eq.~\eqref{eq_gravity} allows us to retain (in almost intact form) all the empirical knowledge that has accumulated in the econometrics literature since Jan Tinbergen's introduction of the GM.
An important difference, however, is that in our model the trade volumes will be drawn from a different probability distribution.
 
The second lesson we have learnt is that, in analogy with Eq.~\eqref{eq_gravityF1}, Eq.~\eqref{eq:fermi} should be generalized to allow for both dyadic ($\vec{D}_{ij}$) and node-specific ($\vec{n}_i$) factors as follows:
\begin{equation}
p_{ij}=\langle a_{ij}\rangle=\frac{G_{\vec{\psi}}\,(\vec{n}_i,\vec{n}_j,\vec{D}_{ij})}{1+G_{\vec{\psi}}\,(\vec{n}_i,\vec{n}_j,\vec{D}_{ij})}\qquad G> 0,
\label{eq:fermiG}
\end{equation}
where a crucial requirement is that $G$ can in general be different from $F$ in Eq.~\eqref{eq_gravityF1} and, correspondingly, the vector $\vec{\psi}$ of parameters can be different from $\vec{\phi}$. 
Note that, since $p_{ij}$ is monotonic in $G$, the above expression is entirely general, i.e. we have put no restriction on the functional form of $p_{ij}$.
It is also worth noticing that the explanatory factors used in Eqs.~\eqref{eq_gravityF1} and~\eqref{eq:fermiG} need not coincide.
However, to avoid using different symbols for the arguments of the two functions, we adopt the convention that $\vec{D}_{ij}$ and $\vec{n}_i$ denote the sets of all factors used as arguments of either $F$ or $G$, and that these functions can have flat (i.e. no) dependence on some of their arguments.
For instance, Eq.~\eqref{eq:fermiG} reduces to Eq.~\eqref{eq:fermi} by setting $\vec{n}_i=\textrm{GDP}_i$ and assuming flat dependence on $\vec{D}_{ij}$, or it reduces to the hyperbolic model in Ref.~\cite{worldtradeatlas} by setting $\vec{D}_{ij}$ equal to the hyperbolic distance and assuming flat dependence on $\vec{n}_i$.

We want our model to produce both Eq.~\eqref{eq_gravityF1} as the desired (gravity-like) conditional expectation for link weights and Eq.~\eqref{eq:fermiG} as a realistic expected topology.
To do so, we introduce the full probability $P(\mathbf{W})$ that the model produces a weighted network specified by the $N\times N$ matrix $\mathbf{W}$ with entries $(w_{ij})$. 
We are free to choose whether $w_{ij}$ takes non-negative integer values (in which case $P(\mathbf{W})$ is a multivariate Probability Mass Function, or PMF) or non-negative real values (in which case $P(\mathbf{W})$ is a multivariate Probability Density Function, or PDF).
The distribution $P(\mathbf{W})$ is the key quantity that fully specifies the model and determines both the topology and the link weights of the ITN.
From $P(\mathbf{W})$, focusing on a single pair $i,j$ of nodes and integrating out all other pairs, we can define the dyadic distribution $q_{ij}(w)$ indicating the probability (mass or density) that $w_{ij}$ takes the particular value $w$.
Note that the event $w_{ij}>0$ indicates the presence of a trade link (i.e. $a_{ij}=1$). 
By contrast, the event $w_{ij}=0$ indicates the absence of a trade link (i.e. $a_{ij}=0$) but is also included as a possible outcome in $q_{ij}(w)$.
The normalization condition is therefore $\sum_{w\ge 0}q_{ij}(w)=1$ (for integer weights) or $\int_{w\ge 0} dw~q_{ij}(w)=1$ (for continuous weights, in which case we anticipate that $q_{ij}(w)$ will have a delta-like point mass at $w=0$) for all $i,j$.
Note that we are \emph{not} assuming independence of the trade volumes $w_{ij}$ and $w_{kl}$ between two distinct country pairs, or equivalently the factorization of $P(\mathbf{W})$ into the product $\prod_{i,j}q_{ij}(w_{ij})$ of dyadic probabilities. 
However, we will later find that the desired model has precisely this independence property.
Importantly, unlike in the traditional GM, in our approach dyadic independence is a consequence and not a postulate.

We now look for the form of $q_{ij}(w)$ that enforces both Eqs.~\eqref{eq_gravityF1} and~\eqref{eq:fermiG}. 
Let us consider the latter first. 
In terms of $q_{ij}(w)$, the probability $p_{ij}$ that $i$ and $j$ are connected (irrespective of the volume of trade) is given by the complement of the probability $q_{ij}(0)$ that they are not connected, i.e.
\begin{equation}
p_{ij}=1-q_{ij}(0)=\left\{\begin{array}{ll}\sum_{w> 0}q_{ij}(w)~&\textrm{(integer)}\\
\int_{w>0}q_{ij}(w)~dw~&\textrm{(real)}\end{array}\right.
\label{eq:1-q}
\end{equation}
where, for real-valued weights, $q_{ij}(0)$ denotes the point mass, i.e. the magnitude of the delta-like probability density function $q_{ij}(w)$, at $w=0$.
Imposing that Eq.~\eqref{eq:1-q} has the form dictated by Eq.~\eqref{eq:fermiG} leads to the following unique choice for $q_{ij}(0)$:
\begin{equation}
q_{ij}(0)=\frac{1}{1+G_{\vec{\psi}}\,(\vec{n}_i,\vec{n}_j,\vec{D}_{ij})}\qquad G> 0.
\label{eq:limit1}
\end{equation}
We now relate $q_{ij}(w)$ to Eq.~\eqref{eq_gravityF1} in a similar manner. 
The expected trade volume, irrespective of whether a link exists, is
\begin{equation}
\langle w_{ij}\rangle\equiv\left\{\begin{array}{ll}\sum_{w> 0}w~q_{ij}(w)&\textrm{(integer)}\\
\int_{w>0}w~q_{ij}(w)~dw~&\textrm{(real)}\end{array}\right.
\label{eq:expw}
\end{equation}
(note that the event $w=0$ does not contribute to the above quantity).
On the other hand the \emph{conditional} probability that $w_{ij}$ equals $w$, given that the link is realized ($w>0$), is 
\begin{equation}
q_{ij}(w|a_{ij}=1)=\left\{\begin{array}{ll}\frac{q_{ij}(w)}{p_{ij}}& w>0\\0&w=0\end{array}\right.
\end{equation}
and its expected value gives the conditional expectation of the link weight, given that the link exists:
\begin{equation}
\langle w_{ij}|a_{ij}=1\rangle=\frac{\langle w_{ij}\rangle}{p_{ij}}.
\label{eq:condw}
\end{equation}

Setting Eq.~\eqref{eq:condw} equal to Eq.~\eqref{eq_gravityF1} leads to
\begin{equation}
\langle w_{ij}\rangle=\frac{F_{\vec{\phi}}\,(\vec{n}_i,\vec{n}_j,\vec{D}_{ij})~G_{\vec{\psi}}\,(\vec{n}_i,\vec{n}_j,\vec{D}_{ij})}{1+G_{\vec{\psi}}(\vec{n}_i,\vec{n}_j,\vec{D}_{ij})}\qquad F,G> 0.
\label{eq:limit2}
\end{equation}

Equation~\eqref{eq:limit2} carries an important message. 
It reveals that, while a superficial inspection of Eq.~\eqref{eq:expw} might suggest that the expected trade volume $\langle w_{ij}\rangle$ is independent of the topology of the ITN, i.e. on $q_{ij}(0)$ or equivalently $G$, this is actually not the case.
In fact, $q_{ij}(0)$ is coupled to the other values $q_{ij}(w)$ (with $w>0$) through the normalization condition manifest in Eq.~\eqref{eq:1-q}. 
This necessarily implies that \emph{the topology of the ITN must have an immediate effect on the expected volume of trade between any two countries.} This effect is rigorously quantified in Eq.~\eqref{eq:limit2}, which shows that $\langle w_{ij}\rangle$ depends on both  $F$ and $G$. 
This result confirms the inconsistency of the traditional GM defined in terms of Eq.~\eqref{eq_gravity} and of any of its extensions of the form given by Eq.~\eqref{eq_gravityF}.
By contrast, \emph{the expected topology of the ITN is independent of the expected volumes of trade}, since $p_{ij}$ depends on $G$ but not on $F$.
This simple but, to the best of our knowledge, previously unrecognized result highlights a nontrivial
asymmetry between weights and topology in the ITN and, by extension, in any (economic) network described by our generic expressions involving $F$ and $G$.
This basic finding provides a natural explanation for the aforementioned empirical observation that the topology of the ITN and several other networks can be satisfactorily reconstructed from aggregate local constraints~\cite{Squartini-random1,ECM}, while the same result does not hold for the weighted structure of the same network(s)~\cite{Squartini-random2,Squartini3}, unless topological information is explicitly included as an additional constraint~\cite{ECM,ECM2}.

\subsection{Maximum entropy construction}
Equations~\eqref{eq:limit1} and~\eqref{eq:limit2} fix two important properties we require for $q_{ij}(w)$ and ultimately $P(\mathbf{W})$, but they do not specify these probability distributions uniquely.
To do so, we invoke the Maximum-Entropy Principle to ensure that the functional form of $P(\mathbf{W})$ is maximally random, given the desired constraints. 
As well known, this procedure is guaranteed to lead to the least biased inference, i.e. to introduce no unjustified `hidden' assumption in picking a specific form of $P(\mathbf{W})$~\cite{mybook,mynaturereviews}.
In applying this method we will focus primarily on the case of integer-valued link weights, since this requirement matches the datasets in our analysis. The case of real-valued link weights is treated in the Appendix and the corresponding key results are briefly reported at the end of this Section.

We look for the form of $P(\mathbf{W})$ that maximizes the entropy functional 
\begin{equation}
S[P]=-\sum_{\mathbf{W}}P(\mathbf{W})\ln P(\mathbf{W})\label{eq:entropy}
\end{equation}
(where the sum extends over all weighted graphs with $N$ nodes, non-negative integer link weights, and such that $w_{ii}=0$ for all $i$) subject to the constraints specified by Eqs.~\eqref{eq:limit1} and~\eqref{eq:limit2}. 
Since Eq.~\eqref{eq:limit1} is equivalent to Eq.~\eqref{eq:fermiG}, we select $\langle a_{ij}\rangle$ and $\langle w_{ij}\rangle$ (for all pairs $i\ne j$) as the two sets of constraints specifying our model.
In this way, if we introduce $\alpha_{ij}$ and $\beta_{ij}$ as the (real-valued) Lagrange multipliers required to enforce the expected value of $a_{ij}=\Theta(w_{ij})$ and $w_{ij}$ respectively (where $\Theta(x)=1$ if $x>0$ and $\Theta(x)=0$ otherwise), then the maximum-entropy problem becomes equivalent to one solved exactly in Ref.~\cite{bosefermi}. 
There, it was shown that upon introducing the  so-called \emph{Hamiltonian}
\begin{equation}
H(\mathbf{W})=\sum_{i,j}\big[ \alpha_{ij} \Theta(w_{ij})+\beta_{ij} w_{ij}\big],
\label{eq:H}
\end{equation}
(representing a linear combination of the quantities whose expected value is being constrained) and the \emph{partition function} $Z=\sum_{\mathbf{W}}e^{-H(\mathbf{W})}$, the maximum-entropy probability $P^*(\mathbf{W})$ with constraints $\langle a_{ij}\rangle$ and $\langle w_{ij}\rangle$ is found to be
\begin{equation}
P^*(\mathbf{W})=\frac{e^{-H(\mathbf{W})}}{Z}=\prod_{i,j}q^*_{ij}(w_{ij}),
\label{probP}
\end{equation}
where, given $x_{ij}\equiv e^{-\alpha_{ij}}\in(0,+\infty)$ and $y_{ij}\equiv e^{-\beta_{ij}}\in(0,1)$,
\begin{equation}
q^*_{ij}(w)\equiv \frac{x_{ij}^{\Theta(w)} y_{ij}^{w}\left( 1-y_{ij}\right)}{1-y_{ij}+x_{ij}y_{ij}},\qquad w\ge 0
\label{probQ}
\end{equation}
is the resulting (maximum-entropy) probability that the link from node $i$ to node $j$ carries a weight $w$. 
This probability is called the Bose-Fermi distribution, as it unifies the Bose-Einstein and Fermi-Dirac distributions encountered in quantum statistical physics~\cite{bosefermi}.
We stress again that all our formulas apply to both directed and undirected representations of the network and, correspondingly, the sums and products over $i,j$ should be interpreted as $i\ne j$ in the directed case (where the pairs $i,j$ and $j,i$ are different) and as $i<j$ in the undirected one (where the pair $i,j$ is the same as the pair $j,i$).
As we had anticipated, the factorization of $P^*(\mathbf{W})$ in terms of products of $q^*_{ij}(w)$ shows that, for this particular choice of the constraints, pairs of nodes turn out to be statistically independent as in the standard GM approach, even if we have not assumed this independence as a postulate in our approach.

Importantly, while the constraints used in the maximum-entropy models of the ITN considered so far in the literature are observed topological properties (e.g. the degrees and/or the strengths of nodes), the constraints considered here are economically-driven expectations, namely Eqs.~\eqref{eq:fermiG} and~\eqref{eq:limit2}. 
This key step allows us to reconcile  macroeconomic and network approaches within a generalized framework, and represents an important difference with respect to previous models.
In particular, we use Eqs.~\eqref{eq:1-q},~\eqref{eq:expw} and~\eqref{eq:condw} to express $p_{ij}$, $\langle w_{ij}\rangle$ and $\langle w_{ij}|a_{ij}=1\rangle$ in terms of $x_{ij}$ and $y_{ij}$~\cite{bosefermi}:
\begin{eqnarray}
p_{ij}&=&1-q^*_{ij}(0)=\frac{x_{ij}y_{ij}}{1-y_{ij}+x_{ij}y_{ij}},\label{eq_ppp*}\\
\langle w_{ij}\rangle&=&\sum_{w> 0}w~q^*_{ij}(w)=\frac{p_{ij}}{1-y_{ij}},\label{eq_w*}\\
\langle w_{ij}|a_{ij}=1\rangle&=& \frac{\langle w_{ij}\rangle}{p_{ij}}=\frac{1}{1-y_{ij}}.
\label{eq_ww*}
\end{eqnarray}
The above expressions allow us to rewrite Eq.~\eqref{probQ} as
\begin{equation}
q^*_{ij}(w)=\left\{\begin{array}{ll}
1-p_{ij}&w=0,\\
p_{ij}~y_{ij}^{w-1}(1-y_{ij})~& w> 0.\end{array}\right.
\label{probQ2}
\end{equation}

Now, equating Eq.~\eqref{eq_ppp*} to Eq.~\eqref{eq:fermiG} and Eq.~\eqref{eq_w*} to Eq.~\eqref{eq:limit2} (or, equivalently, Eq.~\eqref{eq_ww*} to Eq.~\eqref{eq_gravityF1}) allows us to find the values of $x_{ij}$ and $y_{ij}$ solving the original problem:
\begin{eqnarray}
x_{ij}&=&\frac{G_{\vec{\psi}}\,(\vec{n}_i,\vec{n}_j,\vec{D}_{ij})}{F_{\vec{\phi}}\,(\vec{n}_i,\vec{n}_j,\vec{D}_{ij})-1},\label{eq_x}\\
y_{ij} &=& \frac{F_{\vec{\phi}}\,(\vec{n}_i,\vec{n}_j,\vec{D}_{ij})-1}{F_{\vec{\phi}}\,(\vec{n}_i,\vec{n}_j,\vec{D}_{ij})}.\label{eq_y}
\end{eqnarray}
Inserting Eqs.~\eqref{eq_x} and~\eqref{eq_y} into Eq.~\eqref{probQ2}, we finally get the explicit probability $q^*_{ij}(w)$ of any two countries trading a volume $w$, as a function of any choice of the factors $\vec{n}_i$ and $\vec{D}_{ij}$.
In terms of conditional probabilities, the model becomes extremely simple: establishing a link from country $i$ to country $j$ is a Bernoulli trial with success probability $p_{ij}$ given by Eq.~\eqref{eq:fermiG}; if realized, this link acquires a weight $w$ with probability
\begin{equation}
q^*_{ij}(w|a_{ij}=1)=\left\{\begin{array}{ll}
0&w=0,\\
\frac{\left[F_{\vec{\phi}}\,(\vec{n}_i,\vec{n}_j,\vec{D}_{ij})-1\right]^{w-1}}{\left[F_{\vec{\phi}}\,(\vec{n}_i,\vec{n}_j,\vec{D}_{ij})\right]^{w}}& w> 0,\end{array}\right.
\label{probQ**}
\end{equation}
which is a geometric distribution representing the chance of ${w-1}$ consecutive successes, each with  probability $y_{ij}$, followed by a failure with  probability ${1-y_{ij}}$.
The above result provides an insightful interpretation of the realized volumes in the model in terms of processes of link establishment and link reinforcement (see Discussion). 

\subsection{Maximum-Likelihood parameter estimation}
We now take an econometric perspective and discuss how the model parameters can be chosen to optimally fit a specific empirical instance of the network.
To this end, we use the Maximum Likelihood (ML) principle applied to network models~\cite{Unbiased}.
If $\mathbf{W}^*$ denotes the weight matrix (with entries $w^*_{ij}$) of the empirical network, our model generates this particular matrix with probability $P^*(\mathbf{W}^*)$. 
We therefore define the log-likelihood function as
\begin{equation}
\mathcal{L}(\vec{\phi},\vec{\psi})=\ln P^*(\mathbf{W}^*)=\sum_{i,j}\ln\frac{\big(G_{\vec{\psi}}\big)^{a^*_{ij}}\big(F_{\vec{\phi}}-1\big)^{w^*_{ij}-a^*_{ij}}}{\big(1+G_{\vec{\psi}}\big)\big(F_{\vec{\phi}}\big)^{w^*_{ij}}}\nonumber
\end{equation}
(where we have dropped the dependence of $F$ and $G$ on $\vec{n}_i$, $\vec{n}_j$, $\vec{D}_{ij}$)
and look for the parameter values $\vec{\phi}^*,\vec{\psi}^*$ that maximize $\mathcal{L}(\vec{\phi},\vec{\psi})$ by requiring that all the first derivatives with respect to $\vec{\phi}$ and $\vec{\psi}$ vanish simultaneously:
\begin{eqnarray}
\vec{\nabla}_{\vec{\phi}}\,\mathcal{L}(\vec{\phi},\vec{\psi})&=&\sum_{i,j}\left[\frac{w^*_{ij}-a^*_{ij}}{F_{\vec{\phi}}-1}-\frac{w^*_{ij}}{F_{\vec{\phi}}}\right]\vec{\nabla}_{\vec{\phi}}\,F_{\vec{\phi}}
=\vec{0}\label{eq:maxphi}\\
\vec{\nabla}_{\vec{\psi}}\,\mathcal{L}(\vec{\phi},\vec{\psi})&=&\sum_{i,j}\left[\frac{a^*_{ij}}{G_{\vec{\psi}}}-\frac{1}{1+G_{\vec{\psi}}}\right]\vec{\nabla}_{\vec{\psi}}\,G_{\vec{\psi}}
=\vec{0}.\label{eq:maxpsi}
\end{eqnarray}
For probability distributions belonging to the exponential family, i.e. in the form given by Eq.~\eqref{probP} like the one we are considering, the second derivatives of the log-likelihood coincide with (minus) the covariances between the constraints included in the Hamiltonian defined in Eq.~\eqref{eq:H} (see for instance~\cite{covariance,combinatorics}). Since covariance matrices are positive-semidefinite (and actually positive-definite if the chosen constraints are linearly independent, i.e. non-redundant), $\mathcal{L}(\vec{\phi}^*,\vec{\psi}^*)$ is indeed a (global, in the positive-definite case) maximum for $\mathcal{L}(\vec{\phi},\vec{\psi})$, ensuring that the solution $(\vec{\phi}^*,\vec{\psi}^*)$ to Eqs.~\eqref{eq:maxphi} and~\eqref{eq:maxpsi} yields the optimal parameter values in our model.
Selecting these values into Eqs.~\eqref{eq_x} and~\eqref{eq_y} yields the values $x_{ij}^*$ and $y_{ij}^*$ that, when inserted into Eq.~\eqref{probQ}, fully specify the model.

The above expressions, which are valid for \emph{any} specification of the EGM, show that the estimation of the parameter $\vec{\phi}$ nicely separates from that of $\vec{\psi}$.
This result solves, in a single step, two major problems encountered in previous econometric approaches: on one hand, in most alternative models the estimation of the parameters determining the expected weights is badly affected by the presence of the zeroes; on the other hand, the expected number of zeroes may paradoxically depend on the (arbitrary) units of measure for the weights. For instance, if $q_{ij}(w)$ is a Poisson distribution as in zero-inflated GMs~\cite{SantosSilva&Tenreyro_2006,Linders&deGroot_2006,Giorgio_gravity}, then its only parameter (the mean) determines both the magnitude of link weights and the connection probability $p_{ij}$. 
As the monetary units in the data are changed arbitrarily (e.g. from dollars to thousands of dollars), so will the estimated mean and the resulting expected number of zeroes. 
By contrast, in our model 
the monetary units affect $\vec{\phi}^*$ but not $\vec{\psi}^*$ (hence $F$ as they should, but not $G$).

\subsection{Real-valued trade flows}
The above results can be adapted in a straightforward, although more technical, fashion to the case when link weights are assumed to take non-negative real values.
The entire derivation is reported the Appendix. For brevity, here we only report the main results.

In the real-valued case, $P^*(\mathbf{W})$ is a multivariate PDF (rather than a PMF) and we look for its form by maximizing a modified version of the entropy functional $S[P]$, under the same constraints on $\langle a_{ij}\rangle$ and $\langle w_{ij}\rangle$ (for all pairs $i\ne j$) used above and still given by Eqs.~\eqref{eq:fermiG} and~\eqref{eq:limit2}.
The result is again of the factorized form given by Eq.~\eqref{probP}, where the Hamiltonian $H(\mathbf{W})$ is still the one defined in Eq.~\eqref{probP} while the partition function $Z$ is different and the resulting expression for $q^*_{ij}(w)$ is
\begin{equation}
q^*_{ij}(w)=\delta(w)(1-p_{ij})+\Theta(w) \displaystyle{p_{ij}\frac{e^{-w/F_{\vec{\phi}}\,(\vec{n}_i,\vec{n}_j,\vec{D}_{ij})}~}{F_{\vec{\phi}}\,(\vec{n}_i,\vec{n}_j,\vec{D}_{ij})}},
\label{eq:delta3}
\end{equation}
where $\delta(w)$ is the Dirac delta function and $p_{ij}$ is still given by Eq.~\eqref{eq:fermiG}.

The above expression shows that $q^*_{ij}(w)$ has now a point mass of magnitude $q^*_{ij}(0)=1-p_{ij}$ at $w=0$, followed by a purely exponential probability density for $w>0$.
By design, the above PDF still produces the desired conditional expected trade volume $\langle w_{ij}|a_{ij}=1\rangle$, connection probability $p_{ij}$ and unconditional expected trade volume $\langle w_{ij}\rangle$ given by Eqs.~\eqref{eq_gravityF1},~\eqref{eq:fermiG} and~\eqref{eq:limit2} respectively.
Establishing a link from country $i$ to country $j$ is still a Bernoulli trial with success probability $p_{ij}$ given by Eq.~\eqref{eq:fermiG}; if realized, this link acquires a weight $w$ with conditional probability density
\begin{equation}
q^*_{ij}(w|a_{ij}=1)=\left\{\begin{array}{ll}
0&w=0,\\
\frac{e^{-w/F_{\vec{\phi}}\,(\vec{n}_i,\vec{n}_j,\vec{D}_{ij})}}{F_{\vec{\phi}}\,(\vec{n}_i,\vec{n}_j,\vec{D}_{ij})}& w> 0,\end{array}\right.
\label{probQ**real2}
\end{equation}
which is now a purely exponential distribution with the desired (conditional) mean $F_{\vec{\phi}}\,(\vec{n}_i,\vec{n}_j,\vec{D}_{ij})$.

The estimation of the parameters $\vec{\phi}$ and $\vec{\psi}$ can be carried out using the ML principle, via a straightforward recalculation of the log-likelihood $\mathcal{L}(\vec{\phi},\vec{\psi})=\ln P^*(\mathbf{W}^*)$ and a corresponding adaptation of Eqs.~\eqref{eq:maxphi} and~\eqref{eq:maxpsi}.

\section{Empirical analysis}
We can finally test the predictions of our model against empirical international trade data. The datasets are described in the Appendix. Here, it suffices to report that trade volumes are reported in U.S. dollars and are therefore integer-valued. For this reason, throughout our analysis we will adopt the formulas we obtained assuming integer weights. Clearly, the same analysis can be easily repeated for real-valued volumes by using the corresponding formulas we have provided for real weights.

\subsection{Model specification}
We adopt an undirected network description (where the connection between two countries carries a weight equal to the total trade in either direction) to facilitate the definition of the topological properties characterizing the ITN.
Previous work has shown that, given the highly symmetric structure of the ITN, the undirected representation retains all the basic properties of the network~\cite{Garlaschelli,Squartini-random1,Squartini-random2}.

 \begin{figure*}[t]
  \centering
  \includegraphics[width=.9\linewidth]{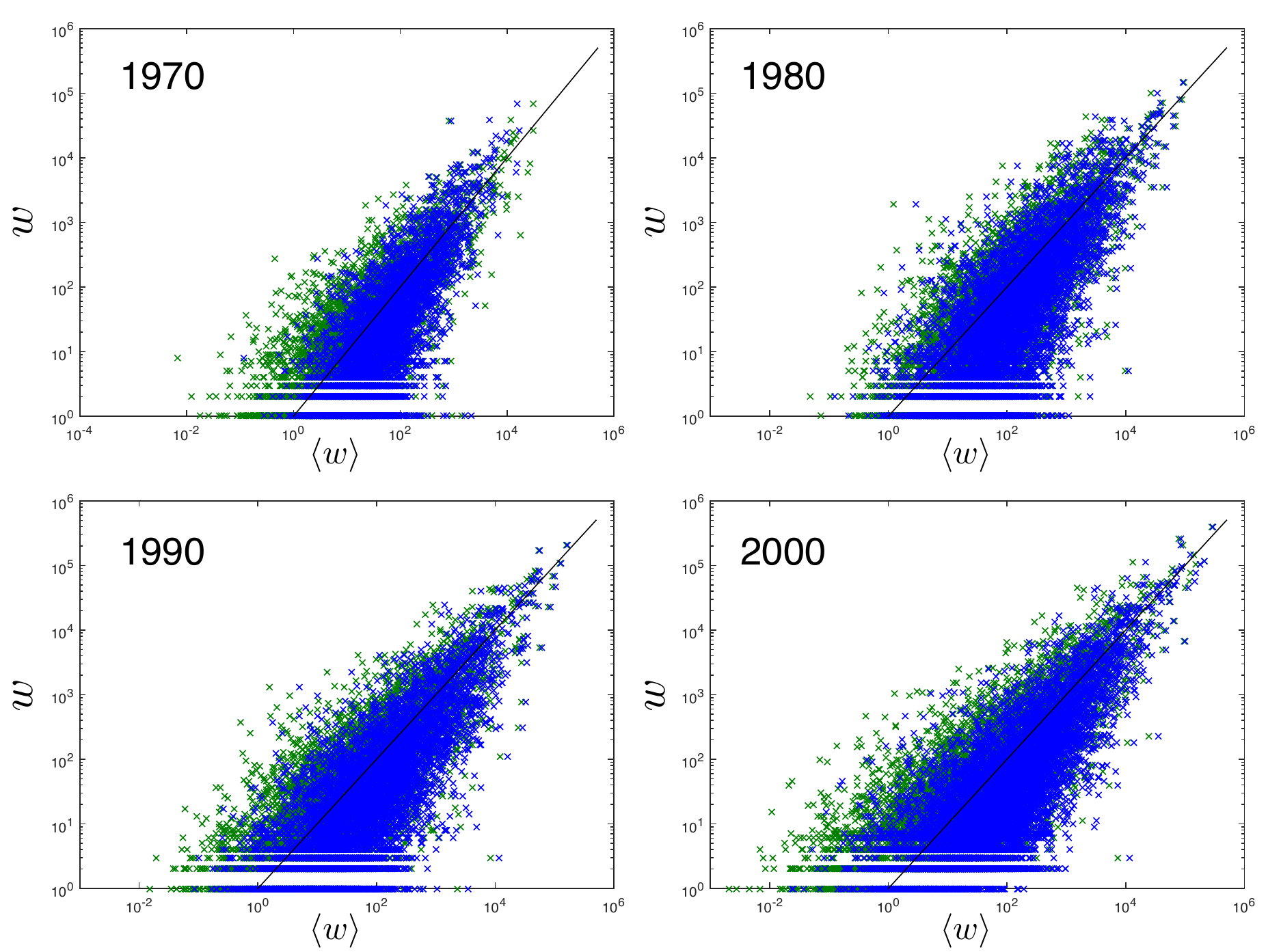}
\caption{{\bf Empirical non-zero trade flows vs. the corresponding expectations under the traditional Gravity Model and the Enhanced Gravity Model.} Log-log plot comparing the empirical volume ($y$-axis) of all non-zero bilateral trade flows in the ITN with the corresponding (conditional) expected volume ($x$-axis) predicted by the Gravity Model defined in Eq.~\eqref{eq_gravity} (green, parameters estimated as reported in Table~\ref{parameters-1}) and by the Enhanced Gravity Model defined in Eqs.~\eqref{eq_gravityF1} and~\eqref{eq:F} (blue, parameters estimated as reported in Table~\ref{parameters-2}). Top left: year 1970, top right: year 1980, bottom left: year 1990, bottom right: year 2000.
The black line is the identity line corresponding to the ideal, perfect match that would be achieved if the empirical weights were exactly equal to their (conditional) expected values, i.e. in complete absence of randomness.}
\label{fig-EGMflows}
\end{figure*}

We choose $F_{\vec{\phi}}(\vec{n}_i,\vec{n}_j,\vec{D}_{ij})$ in such a way that the expected non-zero trade flow $\langle w_{ij}|a_{ij}=1\rangle$ is the same as in the  GM defined by Eq.~\eqref{eq_gravity} (now interpreted as a conditional expectation). This means choosing $\vec{n}_i=\textrm{GDP}_i$, $\vec{D}_{ij}=R_{ij}$, $\vec{\phi}=(c,\alpha,\gamma)$ and 
\begin{equation}
F_{\vec{\phi}}\,(\vec{n}_i,\vec{n}_j,\vec{D}_{ij})=
c~(\textrm{GDP}_i~\textrm{GDP}_j)^{\alpha}~R^{-\gamma}_{ij},
\label{eq:F}
\end{equation}
where we have set $\beta\equiv\alpha$ due to undirectedness. 
Similarly, we choose $G_{\vec{\psi}}(\vec{n}_i,\vec{n}_j,\vec{D}_{ij})$ in such a way that the probability $p_{ij}$ is the same as in the model defined in Eq.~\eqref{eq:fermi}, i.e. $\vec{\psi}=\delta$ and
\begin{equation}
G_{\vec{\psi}}\,(\vec{n}_i,\vec{n}_j,\vec{D}_{ij})=\delta~ \textrm{GDP}_i~\textrm{GDP}_j.
\end{equation}
With the above specification, the expected topology does not depend on any dyadic factor. 
This is the simplest choice that is found to reproduce the topology of the ITN very well~\cite{fitness,Garlaschelli,Caldarelli,Unbiased} and is supported by empirical evidence that dyadic factors like geographic distances~\cite{Distances} and trade agreements~\cite{worldtradeatlas} have a much weaker effect on the purely binary topology of the ITN than on trade volumes.
Of course our formalism has been designed in such a way that we can immediately add dyadic factors, and is therefore much more general.
For instance, we might easily add `hidden' metric distances inferred via an optimal geometric embedding~\cite{worldtradeatlas} (although they would not be identifiable with some empirically measurable, `external' macroeconomic factors like those used elsewhere in our model). 

Given the above model specification, for a given instance $\mathbf{W}^*$ of the empirical network we find the optimal parameter values $c^*$, $\alpha^*$, $\gamma^*$ and $\delta^*$ through the ML conditions given by Eqs.~\eqref{eq:maxphi} and~\eqref{eq:maxpsi}. 
Importantly, Eq.~\eqref{eq:maxpsi} reads in this case $\partial \mathcal{L}/\partial\delta=0$ and yields a value $\delta^*$ that ensures that the expected number of links $\sum_{i,j}p_{ij}=\sum_{i,j}G_{\vec{\psi}}/(1+G_{\vec{\psi}})$ is exactly equal to the empirical number $L^*=\sum_{i,j}a^*_{ij}$, irrespective of the volumes of trade. 
This result, which is equivalent to what is found for the purely binary model defined by Eq.~\eqref{eq:fermi}~\cite{Unbiased}, shows that, unlike the standard GM, our model always generates the correct number of links and, unlike some more complicated variants of the GM, it does so independently of the monetary units chosen for the volumes.

\subsection{Testing the model against real data}
We first test the performance of the EGM in replicating the empirical trade volumes, i.e. the purely local (dyadic) structure of the ITN.
In Fig.~\ref{fig-EGMflows}, superimposed to the previous results for the standard GM given by Eq.~\eqref{eq_gravity} and already shown in Fig.~\ref{fig-GMflows}, the empirical non-zero link weights $w^*_{ij}$ are also compared with their conditional expected value $\langle w_{ij}|a_{ij}=1\rangle$ under the EGM given by Eq.~\eqref{eq:F}. 
As mentioned above, for the EGM the parameters are obtained via the ML principle as prescribed by Eq.~\eqref{eq:maxphi} and their resulting values are reported in Table~\ref{parameters-2}.
As expected, the sets of points generated by the two models largely overlap, confirming that, in terms of trade volumes, the EGM cannot do worse than the GM.
Moreover, the EGM turns out to be more parsimonious than the GM as it achieves a narrower scatter of points while having no dedicated free parameter to tune the variance (as already mentioned, the GM usually assumes that each trade volume is drawn from a certain probability distribution, typically a normal or log-normal one, with mean value given by Eq.~\eqref{eq_gravity} and variance specified by an additional free parameter).

\begin{table}[b]
\centering
\begin{tabular}{|c|c|c|c|c|}
\hline
\multicolumn{5}{|c|}{\bf Enhanced Gravity Model}\\
\hline\hline
\multicolumn{1}{|c|}{Year}&$\delta$ & c & $\alpha$, $\beta$ & $\gamma$ \\ \hline \hline
1970                     & $4.7\cdot10^5$ &$1.0\cdot10^{8}$ &0.67        & 0.78     \\ \hline 
1980                     & $1.1\cdot10^6$  &$9.3\cdot10^{8}$ & 0.77      & 0.75     \\ \hline
1990                     & $1.4\cdot10^6$  &$5.4\cdot10^{9}$ & 0.87     & 0.86     \\ \hline
2000                     & $3.3\cdot10^6$  &$1.7\cdot10^{10}$  & 0.91    &  0.90     \\ \hline
\end{tabular}
\caption{Parameter values for the Enhanced Gravity Model calculated by considering integer link weights (equal to integer multiples of the monetary unit used in the dataset) and carrying out the corresponding ML estimation as prescribed by Eqs.~\eqref{eq:maxphi} and~\eqref{eq:maxpsi}.}
\label{parameters-2}
\end{table}

Importantly, comparing the values of the parameters $\alpha,\beta,\gamma$ reported in Table~\ref{parameters-2} for the EGM with the corresponding values of the same parameters shown previously in Table~\ref{parameters-1} for the GM, we see that the GM yields systematically larger parameter values (especially so for $\alpha,\beta$). This means that, with respect to the EGM, the GM overestimates the effects of both GDP and geographic distance, and this is especially true for the GDP.
This is due to the fact that the EGM is used to explain not only the volume of realized trade flows, but also their existence, and has separate functions ($F$ and $G$) with possibly overlapping sets of explanatory factors (GDP is the common element in this case) but in any case distinct sets of parameters ($\alpha,\beta,\gamma$ on one hand and $\delta$ on the other), to take these two aspects into account.
The effects of GDP and distance captured by the parameters $\alpha,\beta,\gamma$ are only those conditional on a link being created, while discounting the effects of link creation itself via the parameter $\delta$. 
Note that $\alpha,\beta,\gamma$ and $\delta$ are all found to be monotonically increasing over time by the EGM, highlighting a steady increase  of the effects of GDP and distance (even if milder than observed in the GM) and of the density of connections.
In fact, as the network density becomes higher (larger $\delta$ in the EGM), we see a smaller discrepancy between the fitted values of $\alpha,\beta$ in the two models, consistently with the idea that, if all pairs of countries were connected, then both the GM and the EGM would estimate the effects of GDP only through the lens of trade volumes, because the GDP would no longer explain the (fully connected) topology in such an extreme situation.

\begin{figure}[t]
  \centering
  \includegraphics[width=.99\linewidth]{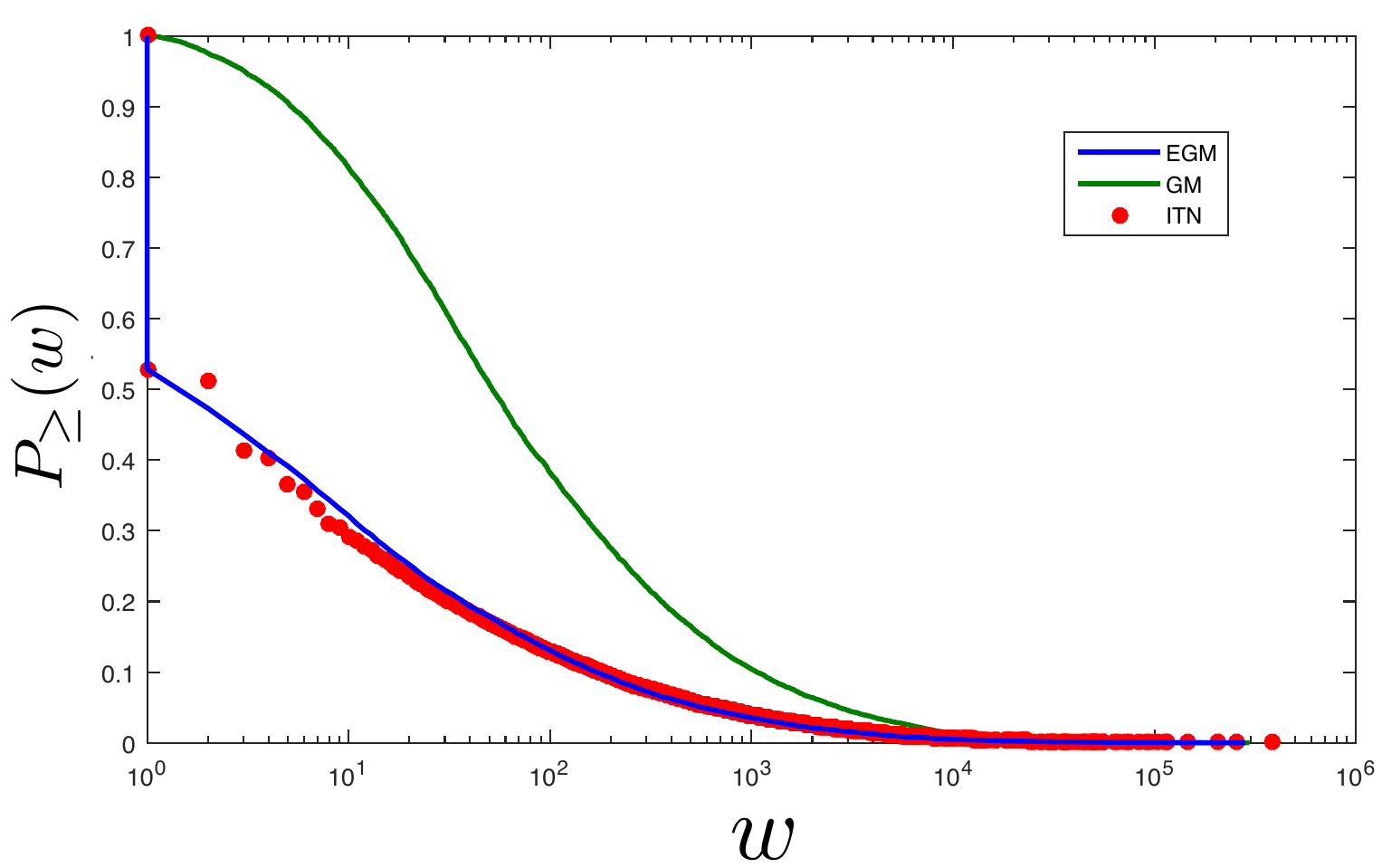}
\caption{{\bf Empirical and model-generated cumulative distributions of trade flows.} Log-linear plot comparing the empirical cumulative distribution of trade flows (normalized in order to include zero flows) in the ITN for the year 2000 (red) with the corresponding distributions obtained using the Gravity Model defined in Eq.~\eqref{eq_gravity} (green, parameters estimated as reported in Table~\ref{parameters-1}) and the Enhanced Gravity Model defined in Eqs.~\eqref{eq_gravityF1} and~\eqref{eq:F} (blue, parameters estimated as reported in Table~\ref{parameters-2}).
Note the discontinuous jump due to the $\approx 47\%$ pairs of unconnected countries in both the empirical and the EGM-generated curves, and the absence of such feature in the GM-generated curve (for which missing links are incorrectly given a positive weight).}
\label{fig-pointmass}
\end{figure}

In order to better understand the differences between the trade volumes predicted by the two models, in Fig.~\ref{fig-pointmass} we plot the cumulative distribution $P_\ge(w)$ counting the fraction of link weights larger than or equal to $w$ in the empirical (red), GM-generated (green) and EGM-generated (blue) networks.
All distributions are normalized as $P_\ge(0)=1$ in order to include zero weights, corresponding to pairs of countries that are \emph{not} connected, in their support. 
Note that $P_{\ge}(w)$ is not simply the integral of $q^*_{ij}(w)$ because the latter is a probability distribution defined for a specific pair of countries, while the former is defined for the entire network and hence determined by the combination of all pair-specific probabilities.
We see that the empirical distribution has a discontinuous jump at $w=1$, as it drops from a value $P_{\ge}(1-\epsilon)=1$ to a value $P_{\ge}(1+\epsilon)\approx 0.53$, where $\epsilon>0$ is arbitrarily small. 
Recalling that link weights take only non-negative integer values in our analysis, this discontinuity indicates that there are roughly $47\%$ pairs of countries that are not connected ($w=0$) in this particular snaphot of the ITN, so that the distribution keeps the value $P_{\ge}(w)=1$ for $w\in[0,1)$ and, as we cross the smallest allowed non-zero weight value ($w=1$), it drops by a value $0.47$ as it no longer `sees' those unconnected pairs.
As bigger weights ($w>1$) are considered, the distribution continues to decrease continuously all the way to $P_{\ge}(+\infty)=0$, indicating that the only discontinuity we see at $w=1$ is actually due to the excess probability mass at zero weights produced by the link-generating process.
Remarkably, the empirical distribution is closely matched by the EGM.
The fact that this model replicates both the location and size of the discontinuity indicates a correctly predicted number of missing trade connections in the ITN topology.
By contrast, the GM predicts a fully connected network, evidenced from the absence of the discontinuity. 
Pair of countries that are unconnected in the real ITN are are unavoidably given a positive weight by the GM and hence misplaced to the right in the distribution, which results in exceedingly large values of the green curve with respect to the other two curves. 
We know that in the EGM the discontinuity is indeed due to the extra point mass at $w=0$ in the expression of $q^*_{ij}(w)$ given by Eqs.~\eqref{probQ} or~\eqref{probQ2}. 
Note that, technically, one can speak of a `discontinuity' only if weights take continuous values. 
This would be possible by replicating our analysis in the case of real-valued weights using the results provided in the Appendix and summarized in Eq.~\eqref{eq:delta3}.
Importantly, in this case the jump in $P_{\ge}(w)$ would be observed precisely at the `true'  value $w=0$, consistently with the genuine delta-like form of $q^*_{ij}(w)$ given by Eq.~\eqref{eq:delta3} (only, it would no longer be possible to show the discontinuity of $P_{\ge}(w)$ along a logarithmic axis and plot the full cumulative distribution). The EGM would again correctly match both location and size of the empirical discontinuity (since $p_{ij}$, hence the expected number of positive weights, is identical in the discrete and continuous versions of the model).
For positive weights, the real-valued EGM would continuously interpolate the discrete points of the integer-valued EGM, because this is a generic property of geometric and exponential distributions with the same expected value. 
So in either specification, the EGM nicely replicates both the empirical distribution of strictly positive link weights and the sharp peak `jumping out' from it, while the GM does not. 

\begin{figure*}[t]
\centering
\includegraphics[width=.9\linewidth]{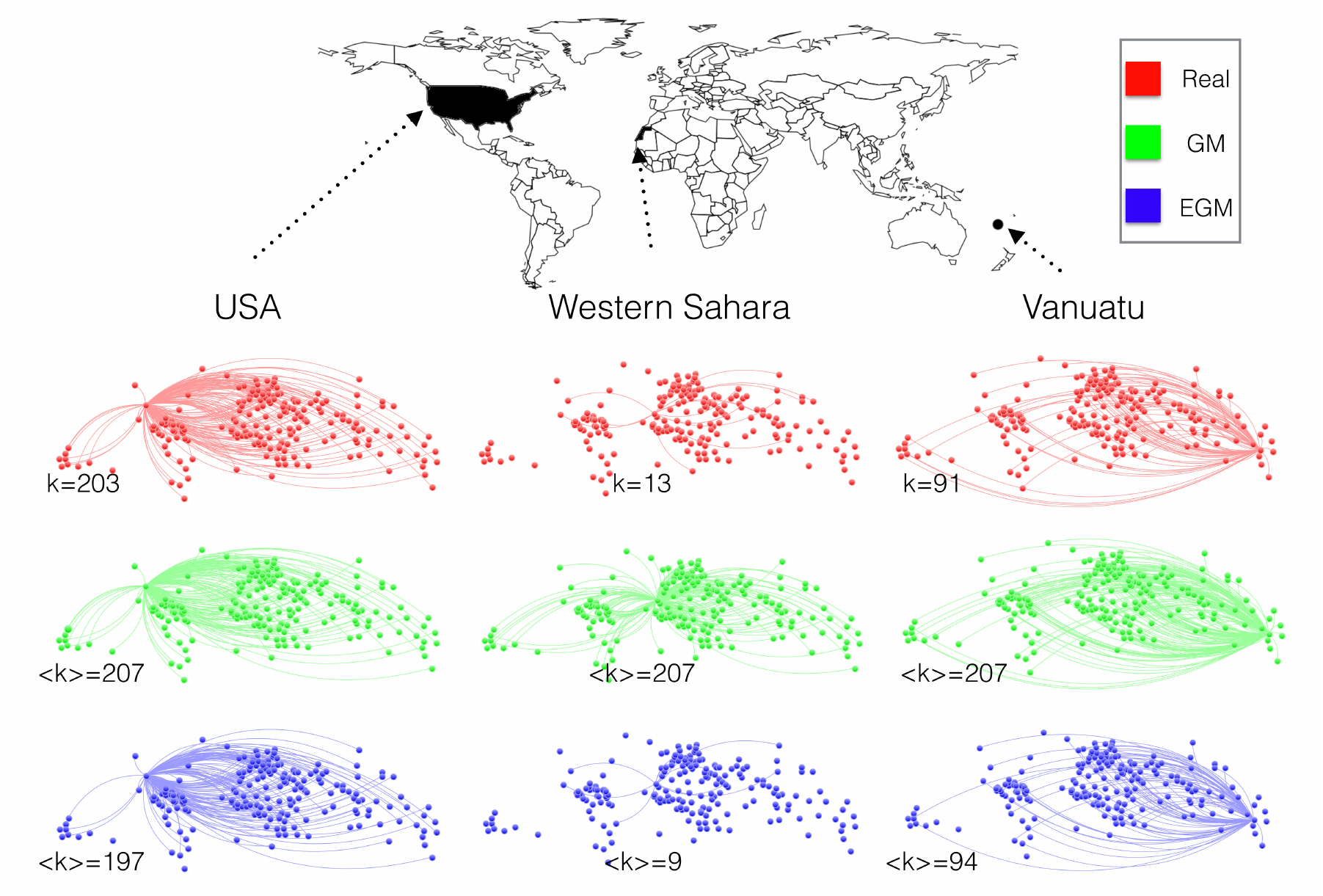}
\caption{{\bf Country-based network configurations for year 2011 in the real ITN (red), the GM (green) and the EGM (blue).} 
For three representative countries, we show the connections to all trade partners in the world.
The total number of countries in the data (see Appendix) is $N=208$.
The three countries are selected on the basis of their empirical degree $k$: the country with maximum degree (USA, $k=203$), the one with minimum degree (Western Sahara, $k=13$) and one with intermediate degree (Vanuatu, $k=91$). 
The GM produces always the maximum possible number ($N-1$) of connections.
By contrast, the EGM produces connections randomly with probability $p_{ij}$, so links change from realization to realization. The expected degree is however independent of the individual realizations and is close to the empirical one for all countries.
We have selected a typical realization that produces a degree equal to the expected degree for each of the three countries.
}
\label{fig-world}
\end{figure*}

We now want to check whether the trade links, besides being predicted in correct number by the EGM, are also placed between the correct pairs of countries by the same model. This means moving the focus of our analysis towards the purely binary, global topology of the ITN.
As a first qualitative illustration setting the stage for this analysis, in Fig.~\ref{fig-world} we show all the trade links of the country with maximum degree (USA), the one with minimum degree (Western Sahara) and one with intermediate degree (Vanuatu). 
We also show the corresponding predictions under the standard GM (where Eq.~\eqref{eq_gravity} is first fitted to the non-zero flows and then extended to all pairs of countries) and the EGM.  
The traditional GM predicts a fully connected network, i.e. an expected degree $\langle k_i\rangle_{GM}=N-1$ for all $i$. This prediction may be accidentally correct for one or a few countries with maximal degree, if such countries turn out to be present in the network (in this case, this does not even happen as the maximum observed degree is $k=203$ for USA), but deteriorates unavoidably and dramatically for other countries as their degree decreases. 
By contrast, the EGM gives an expected degree $\langle k_i\rangle_{EGM}=\sum_{j\ne i}p_{ij}$ (see Appendix) which is in good agreement with the empirical one for the entire range of connectivity. 

\begin{figure*}[t]
\centering
\includegraphics[width=.9\linewidth]{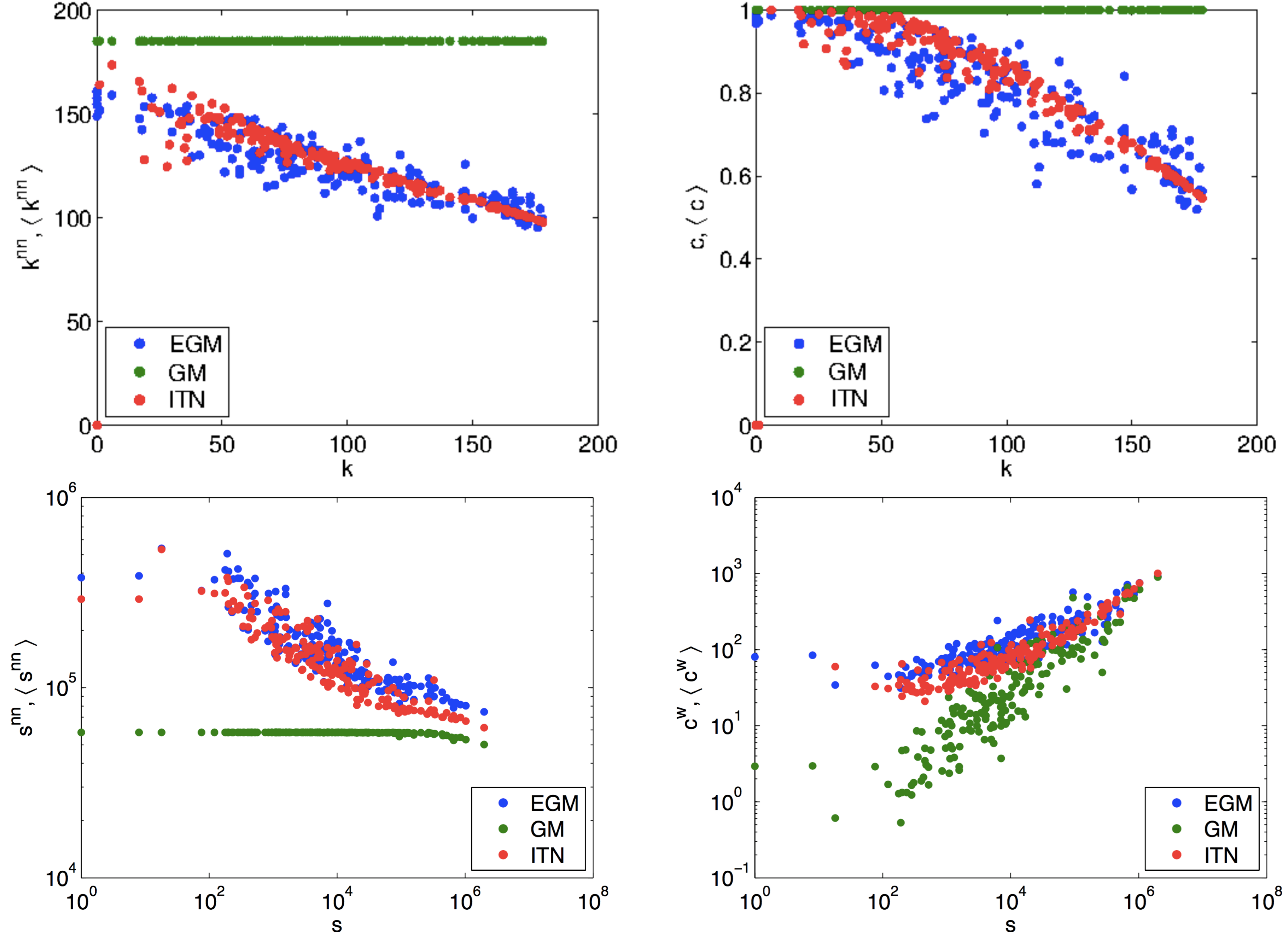} 
\caption{{\bf Network properties in the real ITN (red), the GM (green) and the EGM (blue).} 
Top left: average nearest neighbor degree $k^{nn}_i$ versus degree $k_i$ for all nodes. Top right: clustering coefficient $c_i$ versus degree $k_i$ for all nodes.
Bottom left: average nearest neighbor strength $s_i^{nn}$ versus strength $s_i$ for all nodes. Bottom right: weighted clustering coefficient $c_i^w$ versus strength $s_i$ for all nodes. All results are for the shapshot of the ITN in the year 2000. For all the other years in the analysed sample, we systematically obtained very similar results. See Appendix for information about the data and all definitions of empirical and observed quantities.
}
\label{fig:topology}
\end{figure*}

We now consider higher-order topological properties as a more stringent and quantitative test. 
In the top left panel of Fig.~\ref{fig:topology} we plot the average degree ($k^{nn}_i$) of the trade partners of each country $i$ versus the number of such partners, i.e. the degree ($k_i$) of country $i$ itself. 
Similarly, in the top right panel of Fig.~\ref{fig:topology} we plot the clustering coefficient ($c_i$), i.e. the fraction of trade partners of country $i$ that trade with each other, again versus the number ($k_i$) of such partners. 
The empirical quantities are compared with the expected quantities under the GM and the EGM. The exact expressions for both empirical and expected quantities are provided in Appendix.
The decreasing empirical trends observed in both plots show that the trade partners of poorly connected countries (small $k_i$) are on average highly connected, both to the rest of the world (large $k^{nn}_i$) and among themselves (large $c_i$). 
By contrast, countries that trade with a high-degree country (large $k_i$) are on average poorly connected, both to the rest of the world (small $k^{nn}_i$) and among themselves (small $c_i$). 
For both properties, we find that the EGM is in excellent agreement with the empirical ITN, as opposed to the classical GM which systematically generates nearly constant and much higher values, as a result of predicting a complete network.

Having checked that the EGM does very well in separately replicating both the local link weights and the global topology of the ITN, we now perform a last and most severe test monitoring properties that combine topological and weighted information together (all definitions are again given in the Appendix).
In the bottom left panel of Fig.~\ref{fig:topology} we plot the average strength ($s^{nn}_i$), i.e. the average traded volume, of the trade partners of each country $i$ versus the strength ($s_i$) of country $i$ itself. 
In the bottom right panel, we plot a weighted version of the clustering coefficient ($c^w_i$) of country $i$, again versus the strength ($s_i$) of country $i$. 
The empirical trends are compared with the predictions of the GM and EGM (see Appendix for all definitions). 
These two plots are in some sense the weighted counterparts of the purely binary plots considered above.  
We find that, on average, countries connected to countries with a low trade activity (small $s_i$) trade a lot with the rest of the world (large $s^{nn}_i$) but relatively less so among themselves (small $c^w_i$).
Countries connected to countries with a large volume of trade (large $s_i$) have instead a small trade activity with the rest of the world (small $s^{nn}_i$), but trade relatively strongly with each other (large $c^w_i$). 
Again, we find that both trends are replicated very well by the EGM, while the standard GM fails systematically.

\section{Discussion}
In this paper we have introduced the EGM as a novel, advanced model for the ITN and economic networks in general. 
Phenomenologically, the EGM allows us to reconcile two very different approaches that have remained incompatible so far: on one hand, the traditional GM that is well established in economics and successfully reproduces non-zero trade volumes in terms of GDP and distance but fails in predicting the correct topology~\cite{Giorgio_gravity}; on the other hand, network models that have appeared more recently in the statistical physics literature and have been successful in replicating the topology~\cite{fitness,mynaturereviews} but are more limited in predicting link weights~\cite{Almog}.
To our knowledge, the EGM is the first model that can reliably reproduce the binary and the weighted empirical properties of the ITN simultaneously. 
Just like the standard GM, the RM~\cite{RMmodel} or similar models, the EGM can accomodate additional economic factors in terms of extra dyadic and country-specific properties. 
Yet, it can attribute each of these factors two different roles, by considering its measurable effects on the topology and on the trade volumes separately from each other, although in a combined fashion. For instance, already in the analysis presented here, we have noticed that the EGM uses the GDP in two different ways when explaining the presence and the intensity of links. By discounting the effects of GDP in determining the existence of links from the effects of the same factor in determining the volume of the realized trade connections, the EGM produces different parameter values with respect to the GM. By contrast, the latter lacks this possibility and tends to overestimate the effects of GDP and distances on the measured trade volumes.

The agreement between the EGM and trade data calls for an interpretation of the process generating the network in the model. 
In this respect, we notice that Eqs.~\eqref{probQ} and~\eqref{probQ**} allow us to interpret the realized trade volumes in the EGM as the outcome of two equivalent processes (a serial and a parallel one) of link creation and link reinforcement. 
In the serial process, for a given pair of countries $i,j$ we first establish a trade link of unit weight with success probability $p_{ij}$ and then increment its volume in unit steps, each with success probability $y_{ij}$. After the first failure, we stop the process for the pair of countries under consideration and start it again for a different pair, and so on until all pairs are considered.
In the equivalent parallel process, all pairs of countries simultaneously explore the mutual benefits of trade and engage in a first connection, each with its probability $p_{ij}$. Then, all pairs of nodes for which the previous event has been successful reinforce their existing connection by a unit weight, each with its probability $y_{ij}$. The process stops as soon as there are no more successful events.
In either case, Eq.~\eqref{probQ} gives the resulting probability that the realized volume is $w$.

Importantly, Eq.~\eqref{probQ2} shows that $q^*_{ij}(w)$ is a modified geometric distribution with an extra point mass $q^*_{ij}(0)$ at zero volume, i.e. the first event has a probability $p_{ij}$ which is in general different from the probability $y_{ij}$ of each of the $w-1$ subsequent events required to produce a weight equal to $w$.
This distinguishing property of the Bose-Fermi distribution~\cite{bosefermi} ensures a realistic network formation mechanism where the establishment of a trade connection for the first time is intrinsically different (and therefore associated to a different `cost') from the reinforcement of an already existing trade connection.
This desirable distinction, interpretable for instance in terms of profitability of trade, has been advocated in previous studies~\cite{gravity_book,DeBene2,Linders&deGroot_2006}. 
Here, it is implemented naturally within the maximum-entropy framework via Eq.~\eqref{eq:H}, where the (expected) binary topology is enforced separately from the (expected) link weights.
Notice that the distinction disappears if the parameter $\alpha_{ij}$ in Eq.~\eqref{eq:H} is set to zero, i.e. if the constraint on the expected value of $\Theta(w_{ij})$ (the expected topology) is removed as in the standard GM.
In such a case, $p_{ij}$ becomes equal to $y_{ij}$ (i.e. link creation and link reinforcement become equally likely) and therefore $q_{ij}^*(w)$, not only $q_{ij}^*(w|a_{ij}=1)$, becomes a geometric distribution.
However, this operation would lead to an unrealistically dense network because the expected topology would no longer be controllable separately from the link weights.

Consistently with the fact that trade volumes are typically reported as integer multiples of some indivisible monetary unit (e.g. dollars), the above discussion and most of our analysis has been assuming non-negative integer link weights. 
However we may also take the limit of a vanishing monetary unit, in which case trade volumes become non-negative real numbers and, as we have shown, $q^*_{ij}(w)$ becomes an exponential density with an extra point mass at zero volume as reported in Eq.~\eqref{eq:delta3}, while $q^*_{ij}(w|a_{ij}=1)$ becomes a purely exponential density as shown in Eq.~\eqref{probQ**real2}. 
Crucially, the extra point mass $q^*_{ij}(0)$ ensures that, even in this continuous limit, $p_{ij}$ is unchanged and the expected topology is still described by Eq.~\eqref{eq:fermiG}.
In absence of topological constraints, i.e. if we imposed $\alpha_{ij}=0$, in this real-valued case the network would degenerate to a fully connected graph as in all specifications of the GM with continuous volumes~\cite{fronczakME}. 
This would happen due to the disappearance of the point mass at zero volume, implying that `missing links' become events with zero measure in probability.

Our results may have strong implications both for the theoretical foundations of trade models and for the resulting policy implications. 
It is known that the traditional GM is consistent with a number of (possibly conflicting) micro-founded model specifications \cite{Anderson_1979,Bergstrand_1985,Deardorff_1998, Anderson&vanWincoop_2003}. 
For instance, a gravity-like relation can emerge as the equilibrium outcome of models of trade specialization and monopolistic competition with intra-industry trade \cite{DeBene2,Fratianni_2009}.
The empirical failure of the standard GM  
highlights a previously unrecognized limitation of these micro-founded models, at least in their current form, and indicates the need for an appropriate reformulation that makes these models consistent with the EGM, i.e. with a realistic topology of the ITN.
How policy implications change as the result of such a reformulation of current micro-founded models is an important point to add to the future research agenda.
Research in the field of interbank networks~\cite{myphysicsreports} has shown that, if unrealistically dense networks are assumed, then the outcomes of stress tests typically carried out by central banks to study the propagation of stress among financial institutions are dangerously biased towards a systematic underestimation of systemic risk.
Indeed, running the stress test on a network with the `right' density and topology turns out to be crucial in order to achieve a reliable estimate of risk propagation~\cite{myphysicsreports}.
These results make us confident that, in the field of international economics where the propagation of trade risks is determined by the ITN topology, the EGM may offer a novel benchmark supporting improved theories of trade and refined policy scenarios.



\section*{Appendix}

\subsection*{From integer to real link weights}
If the link weights $w_{ij}$ take non-negative real values instead of non-negative integer values, the probability $P(\mathbf{W})$ has to be interpreted as a PDF rather than a PMF.
We then look for its maximum-entropy functional form $P^*(\mathbf{W})$ by maximizing the following modified version of the entropy introduced in Eq.~\eqref{eq:entropy}:
\begin{equation}
S[P]=-\sum_{\mathbf{A}}\int_{\Theta(\mathbf{W})=\mathbf{A}}d\mathbf{W} P(\mathbf{W})\ln P(\mathbf{W}),\label{eq:entropyreal}
\end{equation}
where the constraints on $\langle a_{ij}\rangle$ and $\langle w_{ij}\rangle$ (for all pairs $i\ne j$) are still given by Eqs.~\eqref{eq:fermiG} and~\eqref{eq:limit2}, and we keep assuming zero-diagonal matrices (no self-loops in the network), i.e. $a_{ii}=w_{ii}=0$ for all $i$.
Note that, in going from Eq.~\eqref{eq:entropy} to Eq.~\eqref{eq:entropyreal}, the summation $\sum_{\mathbf{W}}$ over all $N\times N$ zero-diagonal matrices with non-negative integer entries has been replaced by an integral $\int_{\Theta(\mathbf{W})=\mathbf{A}}d\mathbf{W}$ over all $N\times N$ zero-diagonal matrices with non-negative real entries and such that their binary projection $\Theta(\mathbf{W})$ is a given adjacency matrix $\mathbf{A}$ (i.e. such that $\Theta(w_{ij})=a_{ij}$ for all $i,j$), followed by a discrete sum $\sum_{\mathbf{A}}$ over all such possible binary matrices. 
The resulting integral, written in the combined form $\sum_{\mathbf{A}}\int_{\Theta(\mathbf{W})=\mathbf{A}}d\mathbf{W}$ rather than in the unconstrained form $\int_{\mathbf{W}}d\mathbf{W}$, allows us to treat the binary constraint $\langle a_{ij}\rangle$ more naturally and to recover more general `mixed' (i.e. containing a mixture of a discrete and a continuous part) solutions for $P^*(\mathbf{W})$ that are otherwise inaccessible, as we confirm later.

Since the sets of constraints is the same as in the integer-valued case, we arrive at the same expression for $P^*(\mathbf{W})$ given by \eqref{probP}, where the Hamiltonian $H(\mathbf{W})$ is still given by Eq.~\eqref{eq:H} but, importantly, the partition function $Z$ is now calculated as
\begin{eqnarray}
Z&=&\sum_{\mathbf{A}}\int_{\Theta(\mathbf{W})=\mathbf{A}}d\mathbf{W} ~e^{-H(\mathbf{W})}\nonumber\\
&=&\sum_{\mathbf{A}}\int_{\Theta(\mathbf{W})=\mathbf{A}}d\mathbf{W} ~\prod_{i,j}e^{-\alpha_{ij}\Theta(w_{ij})-\beta_{ij}w_{ij}}\nonumber\\
&=&\prod_{i,j}\sum_{a_{ij}=0,1}\int_{\Theta(w_{ij})=a_{ij}}dw_{ij}~e^{-\alpha_{ij}\Theta(w_{ij})-\beta_{ij}w_{ij}}\nonumber\\
&=&\prod_{i,j}\sum_{a_{ij}=0,1}e^{-\alpha_{ij}a_{ij}}\int_{\Theta(w_{ij})=a_{ij}}dw_{ij}~e^{-\beta_{ij}w_{ij}}\nonumber\\
&=&\prod_{i,j}\left[1+e^{-\alpha_{ij}}\int_{0}^{+\infty}dw_{ij}~e^{-\beta_{ij}w_{ij}}\right]\nonumber\\
&=&\prod_{i,j}\left[1+\frac{x_{ij}}{\beta_{ij}}\right]
\label{eq:partitionreal}
\end{eqnarray}
where we have again used the definition $x_{ij}=e^{-\alpha_{ij}}$, while in this case we find more convenient not to introduce the corresponding transformation $y_{ij}=e^{-\beta_{ij}}$, for reasons that will be clear below.

Inserting Eq.~\eqref{eq:partitionreal} into Eq.~\eqref{probP} yields the following new form of $q^*_{ij}(w)$, replacing the one appearing in Eq.~\eqref{probQ}:
\begin{equation}
q^*_{ij}(w)=\frac{x_{ij}^{\Theta(w_{ij})}\beta_{ij}~e^{-\beta_{ij}w}}{x_{ij}+\beta_{ij}},\quad w\ge0.
\label{probQreal}
\end{equation}
Using Eqs.~\eqref{eq:1-q} and~\eqref{eq:expw}, we can now calculate the connection probability and the (conditional) expected weight as
\begin{eqnarray}
p_{ij}&=&1-q^*_{ij}(0)=\frac{x_{ij}}{x_{ij}+\beta_{ij}},\label{eq_ppp*real}\\
\langle w_{ij}\rangle&=&\int_{w> 0}dw~w~q^*_{ij}(w)=\frac{p_{ij}}{\beta_{ij}},\label{eq_w*real}\\
\langle w_{ij}|a_{ij}=1\rangle&=& \frac{\langle w_{ij}\rangle}{p_{ij}}=\frac{1}{\beta_{ij}}.
\label{eq_ww*real}
\end{eqnarray}
Equations~\eqref{eq_ppp*real},~\eqref{eq_w*real} and~\eqref{eq_ww*real} replace Eqs.~\eqref{eq_ppp*},~\eqref{eq_w*} and~\eqref{eq_ww*} in the case of real-valued link weights.
Inserting these expressions into Eq.~\eqref{probQreal}, we get
\begin{equation}
q^*_{ij}(w)=\left\{\begin{array}{ll}
1-p_{ij}&w=0,\\
p_{ij}~\beta_{ij}~e^{-\beta_{ij}w}~& w> 0,\end{array}\right.
\label{probQ2real}
\end{equation}
which replaces Eq.~\eqref{probQ2} in the real-valued case and shows that $q^*_{ij}(w)$ is now a mixture of a discrete part, characterized by a probability mass of magnitude $q^*_{ij}(0)=1-p_{ij}$ at $w=0$, and a continuous part characterized by an exponential probability density for $w>0$.
If we want to interpret $q^*_{ij}(w)$ uniquely as a PDF throughout its domain (or on the entire real axis), we may rewrite it via the Dirac delta function $\delta(x)$ as
\begin{equation}
q^*_{ij}(w)=\delta(w)(1-p_{ij})+\Theta(w) p_{ij}\beta_{ij}~e^{-\beta_{ij}w},
\label{eq:delta}
\end{equation}
which allows for a fully continuous treatment. For instance, the normalization can be correctly stated as $\int dw~q^*_{ij}(w)=(1-p_{ij})+p_{ij}=1$.
Clearly, the above solution would not be obviously retrieved if we used the unconstrained integral $\int_{\mathbf{W}}d\mathbf{W}$ in Eq.~\eqref{eq:entropyreal}, unless we imposed, \emph{a priori et ad hoc}, the presence of a delta-like spike at zero weight.

In terms of conditional probabilities, we still find that establishing a link from country $i$ to country $j$ is a Bernoulli trial with success probability $p_{ij}$ given by Eq.~\eqref{eq:fermiG} as desired; if realized, this link acquires a weight $w$ with probability density
\begin{equation}
q^*_{ij}(w|a_{ij}=1)=\left\{\begin{array}{ll}
0&w=0,\\
\beta_{ij}~e^{-\beta_{ij}w}~& w> 0,\end{array}\right.
\label{probQ**real}
\end{equation}
which is now a purely exponential distribution with (conditional) mean $\beta_{ij}^{-1}$ as prescribed by Eq.~\eqref{eq_ww*real}.
Now, equating Eq.~\eqref{eq_ppp*real} to Eq.~\eqref{eq:fermiG} and  Eq.~\eqref{eq_ww*real} to Eq.~\eqref{eq_gravityF1} yields the values of $x_{ij}$ and $\beta_{ij}$ solving the original problem:
\begin{eqnarray}
x_{ij}&=&\frac{G_{\vec{\psi}}\,(\vec{n}_i,\vec{n}_j,\vec{D}_{ij})}{F_{\vec{\phi}}\,(\vec{n}_i,\vec{n}_j,\vec{D}_{ij})},\label{eq_xreal}\\
\beta_{ij} &=& \frac{1}{F_{\vec{\phi}}\,(\vec{n}_i,\vec{n}_j,\vec{D}_{ij})}.\label{eq_betareal}
\end{eqnarray}
Note that Eq.~\eqref{eq:limit2} holds in this case as well, as it should because it does not depend on whether link weights are taken to be integer or real.
Inserting Eqs.~\eqref{eq_xreal} and~\eqref{eq_betareal} into Eqs.~\eqref{eq:delta} and~\eqref{probQ**real}, we get the explicit form of $q^*_{ij}(w)$ and $q^*_{ij}(w|a_{ij}=1)$ as a function of the factors $\vec{n}_i$ and $\vec{D}_{ij}$, as reported in the main text in Eqs.~\eqref{eq:delta3} and~\eqref{probQ**real2} respectively.

\subsection*{Data}
We have used international trade and GDP data from the  database curated by Gleditsch~\cite{Gleditsch} for the years 1950, 1960, 1970, 1980, 1990 and 2000. This database includes yearly trade volumes $w_{ij}$ (which we have symmetrized by taking the sum of $w_{ij}+w_{ji}$), yearly GDP values, and the (time-independent) distance matrix $R_{ij}$. The number $N$ of countries increases over time from roughly 85 in 1950 to approximately 200 in 2000. Both GDP and trade data are reported in U.S. dollars and are therefore integer-valued. 
To produce Fig.~\ref{fig-world}, we have used the BACI database~\cite{BACI}, which reports imports and exports between $N=208$ countries in 2011. The BACI data were originally in disaggregated form, where total trade was resolved into 96 different non-overlapping commodity classes. We have aggregated all these commodity classes together, and again symmetrized, to obtain a dataset consistent with the Gleditsch data used for the earlier years. 

\subsection*{Observed network properties}
Given a weighted undirected network with weight matrix $\mathbf{W}$ and adjacency matrix $\mathbf{A}$, with entries related through $a_{ij}=\Theta(w_{ij})$, the \emph{degree} of node $i$ is defined as
\begin{equation}
 k_i=\sum_{j\neq i}a_{ij},
\label{eq:k}
\end{equation}
the \emph{average nearest-neighbor degree} of node $i$ is defined as
\begin{equation}
 k_i^{nn}=\sum_{j\neq i} \frac{a_{ij} k_j}{k_i}=\frac{\sum_{j\neq i}\sum_{k\neq j}a_{ij}a_{jk}}{\sum_{j\neq i}a_{ij}},
\label{eq:knn}
\end{equation}
and the \emph{(binary) clustering coefficient} of node $i$ is defined as
\begin{equation}
 c_i=\frac{\sum_{j\neq i}\sum_{k\neq i,j}a_{ij}a_{jk}a_{ki}}{\sum_{j\neq i}\sum_{k\neq i,j}a_{ij}a_{ki}}.
\label{eq:cbin}
\end{equation}
The \emph{average nearest neighbor strength} of node $i$ is defined as 
\begin{equation}
 s_i^{nn}=\sum_{j\neq i} \frac{a_{ij} s_j}{k_i}=\frac{\sum_{j\neq i}\sum_{k\neq j}a_{ij}w_{jk}}{\sum_{j\neq i}a_{ij}}
\label{eq:snn}
\end{equation}
(where $s_i=\sum_{j\ne i} w_{ij}$ is the \emph{strength} of node $i$) and the \emph{weighted clustering coefficient} of node $i$ is defined as
\begin{equation}
 c_i^{w}=\frac{\sum_{j\neq i}\sum_{k\neq i,j}(w_{ij}w_{jk}w_{ki})^{\frac{1}{3}}}{\sum_{j\neq i}\sum_{k\neq i,j}a_{ij}a_{ki}}.
\label{eq:cwei}
 \end{equation}

\subsection*{Expected network properties}
The expected value (under the EGM) of each of the network properties defined above can be calculated either numerically, by averaging over many network realizations sampled independently from the probability $P^*(\mathbf{W})$ in Eq.~\eqref{probP}, or analytically, using the following approach. First of all, in this model the expected value of all ratios can be approximated by the ratio of the expected values~\cite{ECM,ECM2}.
Secondly, all numerators and denominators involve only products over distinct pairs of nodes, which are statistically independent in the model. Using Eq.~\eqref{probQ}, the expected values of such products can therefore be calculated exactly in terms of $x_{ij}$ and $y_{ij}$ as follows:
\begin{eqnarray}
 \left\langle \sum_{i,j,k,\dots} a_{ij} \cdot a_{jk}\cdot ... \right\rangle&=& \sum_{i,j,k,\dots} \langle a_{ij}\rangle \cdot  \langle a_{jk}\rangle \cdot \langle ... \rangle,\\
\left\langle \sum_{i,j,k,\dots} w_{ij}^{\alpha} \cdot w_{jk}^{\beta}\cdot ... \right\rangle&=& \sum_{i,j,k,\dots} \langle w_{ij}^{\alpha}\rangle \cdot  \langle w_{jk}^{\beta}\rangle \cdot \langle ... \rangle,
\end{eqnarray}
where $\langle a_{ij}\rangle=p_{ij}$, as given by Eq.~\eqref{eq_ppp*}, and 
\begin{equation}
 \langle w_{ij}^{\gamma}\rangle\equiv\sum_{w=0}^{\infty} w^{\gamma} q_{ij} (w)= \frac{x_{ij}(1-y_{ij})\textrm{Li}_{-\gamma}(y_{ij})}{1-y_{ij}+x_{ij}y_{ij}},
 \label{poly}
\end{equation}
$\textrm{Li}_n(z)=\sum_{l=1}^{\infty} \frac{z^l}{l^n}$ denoting the so-called $n-$th polylogarithm of $z$.
From the above two considerations, it follows that the expected properties of all quantities of interest can be approximated with entirely analytical expressions obtained by simply replacing $a_{ij}$ with $p_{ij}$ and $w^\gamma_{ij}$ with $\langle w^\gamma_{ij}\rangle$ in Eqs.~\eqref{eq:k},~\eqref{eq:knn},~\eqref{eq:cbin},~\eqref{eq:snn} and~\eqref{eq:cwei}.
Via $x_{ij}$ and $y_{ij}$, the expected values are ultimately a function of only the GDPs and distances.
In our analysis, after preliminary checking that the analytical expressions matched extremely well with the numerical averages over realizations, we have systematically adopted the analytical approach, which requires no sampling of networks and is therefore extremely efficient.

\section*{Acknowledgements}
AA and DG acknowledge support from the Dutch Econophysics Foundation (Stichting Econophysics, Leiden, the Netherlands). 
This work was also supported by the Netherlands Organization for Scientific Research (NWO/OCW).



\begin{thebibliography}{99}

\bibitem{economicnetworks}
F. Schweitzer, G. Fagiolo, D. Sornette, F. Vega-Redondo, A. Vespignani, D.R. White. {\em Economic networks: The new challenges}, Science {\bf 325(5939)}, 422 (2009).

\bibitem{Kali1} 
R. Kali and J. Reyes,{\em The architecture of globalization: a network approach to international economic integration},
Journal of International Business Studies {\bf38}, 595 (2007).

\bibitem{Kali2} 
R. Kali and J. Reyes, {\em Financial contagion on the international trade network},
Economic Inquiry {\bf 48}, 1072 (2010).

\bibitem{integration}
S. Schiavo, J. Reyes, G. Fagiolo, {\em International trade and financial integration: a weighted network analysis}, Quantitative Finance {\bf 10(4)}, 389-399 (2010).

\bibitem{Spitz}
J. Spitz, T. Kastelle, {\em Gains from Trade: the Impact of International Trade on National Economic Convergence; A Complex Network Analysis Approach}, Eastern Economic Association Annual Meeting, 2010.


\bibitem{myscience}
S. Battiston et al.,
\emph{Complexity theory and financial regulation}, Science 351:6275, 818-819 (2016).

\bibitem{TradeRisk}
P. Klimek, M. Obersteiner and S. Thurner,
{\em Systemic trade risk of critical resources}, Science Advances,
{\bf 1}, e1500522 (2015).


\bibitem{Tinbergen1}
J. Tinbergen, {\em Shaping the World Economy: Suggestions for an International Economic Policy}, 
(The Twentieth Century Fund, New York, 1962).

\bibitem{gravity_book}
P. van Bergeijk, S. Brakman (eds.) {\em The gravity model in international trade} (Cambridge University
Press, Cambridge, 2010).

\bibitem{DeBene2}
L. De Benedictis, D. Taglioni, {\em The gravity model in international trade}, in The Trade Impact of European Union Preferential Policies, pp. 55-89 (Springer Berlin Heidelberg, 2011).

\bibitem{RMmodel}
F. Simini, M.C. Gonz\'alez, A. Maritan, A. Barab\'asi {\em A universal model for mobility and migration patterns}
Nature {\bf 484}, 96–100 (2012).

\bibitem{zipf}
G. K. Zipf, 
\emph{The $P_1 P_2/D$ hypothesis: on the intercity movement of persons},
Am. Sociol. Rev. {\bf 11}, 677-686 (1946).

\bibitem{balcan}
D. Balcan, V. Colizza, B. Gon\c{c}alves, H. Hu, J. J. Ramasco, A. Vespignani,
\emph{Multiscale mobility networks and the spatial spreading of infectious diseases},
Proc. Natl. Acad. Sci. USA {\bf 106}(51), 21484-21489 (2009).

\bibitem{koreaGM}
W.-S. Jung, F. Wang, H. E. Stanley,
\emph{Gravity model in the Korean highway},
EPL {\bf 81}, 48005 (2008).

\bibitem{LambiotteGM}
G. Krings, F. Calabrese, C. Ratti, V. D. Blondel,
\emph{Urban gravity: a model for inter-city telecommunication flows},
J. Stat. Mech. {\bf 2009}, L07003 (2009).

\bibitem{migrationGM}
G. Fagiolo, M. Mastrorillo,
\emph{International migration network: Topology and modeling},
Phys. Rev. E {\bf 88}, 012812 (2013).

\bibitem{ravenstein}
E. G. Ravenstein, 
\emph{The Laws of Migration}, 
Journal of the Royal Statistical Society {\bf 52}(2), 241-305.

\bibitem{Glick&Rose_2001}
R. Glick and A.K. Rose {\em Does a Currency Union affect Trade? The Time Series Evidence},
NBER Working Paper No. 8396 (2001).

\bibitem{Rose&Spiegel_2004}
A. K. Rose, M. M. Spiegel {\em A Gravity Model of Sovereign Lending: Trade, Default, and Credit}
IMF Staff Papers, Vol. 51, Special Issue, International Monetary Fund (2004)

\bibitem{SantosSilva&Tenreyro_2006}
J. Silva and S.Tenreyro {\em The Log of Gravity}
The Review of Economics and Statistics, {\bf 88}, issue 4, pages 641-658 (2006).

\bibitem{Linders&deGroot_2006}
G.J. Linders and H.L.F. de Groot {\em Estimation of the Gravity Equation in the Presence of Zero Flows}
Tinbergen Institute Discussion Paper No. 06-072/3 (2006). 

\bibitem{Giorgio_gravity}
M. Duenas, G. Fagiolo, {\em Modeling the International-Trade Network: A Gravity Approach},
Journal Of Economic Interaction And Coordination {\bf 8} 155-178 (2013).

\bibitem{Giorgio_gravity0}
G. Fagiolo, {\em The international-trade network: gravity equations and topological properties}, Journal of Economic Interaction and Coordination {\bf 5(1)}, 1-25 (2010).

\bibitem{Tinbergen2}
T. Squartini and D. Garlaschelli, {\em Jan Tinbergen's legacy for economic networks: from the gravity model to quantum statistics}, in Econophysics of Agent-Based Models, pp. 161-186 (Springer International Publishing, 2014).

\bibitem{fitness}
D. Garlaschelli, M.I. Loffredo, {\em Fitness-Dependent Toplogical Properties of the World Trade Web}, Phys. Rev. Lett {\bf 93}, 188701 (2004).

\bibitem{Squartini-random1}
T. Squartini, G. Fagiolo, D. Garlaschelli, {\em Randomizing world trade. I. A binary network analysis}, Phys. Rev. E {\bf 84}, 046117 (2011).

\bibitem{Squartini-random2}
T. Squartini, G. Fagiolo, D. Garlaschelli, {\em Randomizing world trade. II. A weighted network analysis}, Phys. Rev. E {\bf 84}, 046118 (2011).

\bibitem{Squartini3}
G. Fagiolo, T. Squartini, D. Garlaschelli, {\em Null Models of Economic Networks: The Case of the World Trade Web}, 
J. Econ. Interac. Coord. {\bf 8}(1), 75-107 (2013).

\bibitem{Serrano} 
A. Serrano and M. Boguna,{\em Topology of the world trade web},
Phys. Rev. E {\bf68},015101(R) (2003).

\bibitem{Garlaschelli} 
D. Garlaschelli and M. Loffredo, {\em Structure and Evolution of the World Trade Network}, 
Physica A {\bf355}, 138 (2005).

\bibitem{Vespignani}
A. Serrano, M. Boguna, and A. Vespignani, {\em Patterns of dominant flows in the world trade web},
J. Econ. Interact. Coord.{\bf 2}, 111 (2007).

\bibitem{Caldarelli}
D. Garlaschelli, T. Di Matteo, T. Aste, G. Caldarelli, and M. Loffredo, {\em Interplay between topology and dynamics in the World Trade Web},
Eur. Phys. J. B {\bf 57}, 1434 (2007).

\bibitem{Fagiolo1}
G. Fagiolo, J. Reyes, S. Schiavo, {\em On the topological properties of the world trade web: A weighted network analysis}, Physica A {\bf 387(15)}, 3868-3873 (2008).

\bibitem{Fagiolo2}
G. Fagiolo, J. Reyes, S. Schiavo, {\em World-trade web: Topological properties, dynamics, and evolution}, Physical Review E {\bf 79(3)}, 036115 (2009).

\bibitem{Fagiolo3}
G. Fagiolo, J. Reyes, S. Schiavo, {\em The evolution of the world trade web: a weighted-network analysis}, Journal of Evolutionary Economics, {\bf 20(4)}, 479-514 (2010).

\bibitem{Colizza}
V. Colizza, A. Flammini, M. A. Serrano, and A. Vespignani, {\em Detecting rich-club ordering in complex networks}, Nature Physics {\bf 2}, 110 - 115 (2006).

\bibitem{Zlatic}
V. Zlatic, G. Bianconi, A. D. Guilera, D. Garlaschelli, F. Rao, and G. Caldarelli,  {\em On the rich-club effect in dense and weighted networks}
Eur. Phys. J. B {\bf 67}, 271-275 (2009).

\bibitem{Unbiased}
D. Garlaschelli, M.I. Loffredo, {\em Maximum Likelihood: Extracting Unbiased Information from Complex Networks}, Phys. Rev. E {\bf 78}, 015101(R) (2008).

\bibitem{fronczakME}
A. Fronczak, P. Fronczak, J.A. Holyst, (2012) `Statistical mechanics of the international trade network', 
{\em Phys. Rev. E}, Vol. 85, pp.056113

\bibitem{ECM}
R. Mastrandrea, T. Squartini, G. Fagiolo, D. Garlaschelli, {\em Enhanced reconstruction of weighted networks from strengths and degrees}, New J. Phys. {\bf 16}, 043022 (2014).

\bibitem{ECM2}
R. Mastrandrea, T. Squartini, G. Fagiolo, D. Garlaschelli, {\em Reconstructing the world trade multiplex: the role of intensive and extensive biases}, Phys. Rev. E  {\bf 90}, 062804 (2014).

\bibitem{Almog}
A. Almog, T. Squartini, and D. Garlaschelli,  {\em A GDP-driven model for the binary and weighted structure of the International Trade Network}, New J. Phys. {\bf17}, 013009 (2015).

\bibitem{mybook}
T. Squartini, D. Garlaschelli, {\em Maximum-Entropy Networks: pattern detection, network reconstruction, and graph combinatorics} (Springer International Publishing AG, SpringerBriefs in Complexity, 2017). 

\bibitem{mynaturereviews}
G. Cimini, T. Squartini, F. Saracco, D. Garlaschelli, A. Gabrielli, G. Caldarelli, {\em The statistical physics of real-world networks}, Nature Reviews Physics {\bf 1}, 58-71 (2019).

\bibitem{myphysicsreports}
T. Squartini, G. Caldarelli, G. Cimini, A. Gabrielli, D. Garlaschelli, {\em Reconstruction methods for networks: the case of economic and financial systems}, Physics Reports {\bf 757}, 1-47 (2018).


\bibitem{LGTM}
T. Deguchi, H. Takayasu, M. Takayasu, {\em Simulation of Gross Domestic Product in International Trade Networks: Linear Gravity Transportation Model}, Proceedings of the International Conference on Social Modelling and Simulation, plus Econophysics Colloquium 2014 (2015).

\bibitem{worldtradeatlas}
G. Garc\'ia-P\'erez, M. Bogu\~{n}\'a, A. Allard, M. Serrano, {\em The hidden hyperbolic geometry of international trade: World Trade Atlas 1870–2013}, Scientific Reports, {\bf 6}, 33441, (2016).

\bibitem{bosefermi}
D. Garlaschelli, M.I. Loffredo, {\em Generalized Bose-Fermi Statistics and Structural Correlations in Weighted Networks}, Phys. Rev. Lett. {\bf 102}, 038701 (2009)

\bibitem{covariance}
D. Garlaschelli, F. den Hollander, A. Roccaverde, {\em Covariance structure behind breaking of ensemble equivalence in random graphs}, Journal of Statistical Physics {\bf 173:3-4}, 644-662 (2018). 

\bibitem{combinatorics}
T. Squartini, D. Garlaschelli, {\em Reconnecting statistical physics and combinatorics beyond ensemble equivalence}, {\tt https://arxiv.org/abs/1710.11422}

\bibitem{Distances}
F. Picciolo, T. Squartini, F. Ruzzenenti, R. Basosi, D. Garlaschelli, {\em The role of distances in the World Trade Web}, Proceedings of the Eighth International Conference on Signal-Image Technology \& Internet-Based Systems (SITIS 2012), pp. 784-792 (edited by IEEE) (2013).

\bibitem{Anderson_1979}
J. E. Anderson, {\em A Theoretical Foundation for the Gravity Equation},
American Economic Review, {\bf69} (1), 106-16 (1979).

\bibitem{Bergstrand_1985}
J. Bergstrand, {\em The Gravity Equation in International Trade: Some Microeconomic Foundations and Empirical Evidence},
The Review of Economics and Statistics, vol. 67, issue 3, pages 474-81 (1985).

\bibitem{Deardorff_1998}
A.V. Deardorff, {\em Determinants of Bilateral Trade:Does Gravity Work in a Neoclassical World?}
NBER Working Paper, No. 5377, (1995).

\bibitem{Anderson&vanWincoop_2003}
J.E. Anderson, E. Wincoop, {\em Gravity with Gravitas: A Solution to the Border Puzzle},
NBER Working Paper, No. 8079, (2001).





\bibitem{Fratianni_2009}
M.Fratianni, {\em Expanding RTAs, trade flows, and the multinational enterprise}, 
Journal of International Business Studies, Vol 40, {\bf 7}, 1206-1227, (2009). 

\bibitem{Gleditsch}
Gleditsch, K. S. (2002). {\em Expanded trade and GDP data}. Journal of Conflict Resolution, 46(5), 712-724.

\bibitem{BACI}
G. Gaulier and S. Zignago, CEPII Working Paper 23 (2010).

\end{thebibliography}
\end{document}